\documentclass[preprint2]{aastex}

\shorttitle{Torus as Wind}
\shortauthors{}
\newcommand{\um}{{\,\mu\rm m}}

\newcommand{\msun}{{\rm\,M_\odot}}

\begin{document}

\bibliographystyle{apj}

\title{Sweeping Away the Mysteries of Dusty Continuous Winds in AGN}

\author{S. K. Keating}
\affil{University of Western Ontario, Department of Physics \& Astronomy; University of Toronto, Department of Astronomy \& Astrophysics}
\email{stephanie.keating@utoronto.ca}

\author{J. E. Everett}
\affil{University of Wisconsin--Madison, Department of Physics}

\author{S. C. Gallagher} \and \author{R. P. Deo}
\affil{University of Western Ontario, Department of Physics \& Astronomy}

% =====================================================
% ABSTRACT
% =====================================================
\begin{abstract}
  An integral part of the Unified Model for Active Galactic Nuclei
  (AGNs) is an axisymmetric obscuring medium, which is commonly
  depicted as a torus of gas and dust surrounding the central
  engine. However, a robust, dynamical model of the torus is required
  in order to understand the fundamental physics of AGNs and interpret
  their observational signatures. Here we explore self-similar, dusty
  disk-winds, driven by both magnetocentrifugal forces and radiation
  pressure, as an explanation for the torus. Using these models, we
  make predictions of AGN infrared (IR) spectral energy distributions
  (SEDs) from $2 - 100 \um$ by varying parameters such as: the viewing
  angle (from $i = 0^{\circ} - 90^{\circ}$); the base column density
  of the wind (from $N_{H,0} = 10^{23} - 10^{25}$ cm$^{-2}$); the
  Eddington ratio (from $L/L_{Edd} = 0.01-0.1$); the black hole mass
  (from $M_{BH} = 10^{8} - 10^{9} \msun$); and the amount of power in
  the input spectrum emitted in the X-ray relative to that emitted in
  the UV/optical (from $\alpha_{ox} = 1.1 - 2.1$). We find that models
  with $N_{H,0} = 10^{25}$ cm$^{-2}$, $L/L_{Edd} = 0.1$, and $M_{BH}
  \geq 10^{8} \msun$ are able to adequately approximate the general
  shape and amount of power expected in the IR as observed in a
  composite of optically luminous Sloan Digital Sky Survey (SDSS)
  quasars.  The effect of varying the relative power coming out in
  X-rays relative to the UV is a change in the emission below
  $\sim5$~\micron\ from the hottest dust grains; this arises from the
  differing contributions to heating and acceleration of UV and X-ray
  photons.  We see mass outflows ranging from $\sim1$--4
  $\msun$~yr$^{-1}$, terminal velocities ranging from $\sim1900$--8000
  km sec$^{-1}$, and kinetic luminosities ranging from
  $\sim1\times10^{42}$--$8 \times 10^{43}$ erg s$^{-1}$.  Further
  development of this model holds promise for using specific features
  of observed IR spectra in AGNs to infer fundamental physical
  parameters of the systems.
  \end{abstract}
%slight re-wording of abstract by JEE

\keywords{galaxies: active --- galaxies: Seyfert --- hydrodynamics --- MHD ---
  infrared: general --- quasars: general --- radiative transfer}

% =====================================================
% INTRODUCTION
% =====================================================
\section{Introduction}\label{Intro}

Active galactic nuclei (AGNs) are often set apart from other
astrophysical objects by their powerful spectral energy distributions
(SEDs) that radiate a substantial amount of power across the
electromagnetic spectrum \citep[e.g.,][]{Elvis1994}. In addition to
being powerful, AGNs are also a motley lot, known for displaying a
wide range of observable characteristics among them. Objects of two
types -- radio-loud and radio-quiet -- have a range of intrinsic
luminosities, and may display either only narrow, or both narrow and
broad spectral emission lines \citep{Antonucci1993}.

The Unified Model \citep[e.g.,][]{Antonucci1993, Urry1995} is a
phenomenological depiction of AGNs that posits that the nature of many
of these differences, particularly the presence or absence of broad
emission lines, can be explained primarily as a function of viewing
angle. Pictured at the core of this model is the ``central engine,"
composed of the supermassive black hole (with a mass from $\sim
10^{6}$--$10^{9} \msun$) and its sub-parsec-scale accretion disk,
where matter is heated as it falls towards the black hole. Another key
feature of the Unified Model is a structure that surrounds the
accretion disk: an axisymmetric, obscuring medium, which covers many
lines of sight to the accretion disk around the central black
hole. This obscuring medium is commonly depicted as a torus of
optically thick dusty gas which can extend up to 100 pc
\citep[e.g.,][]{Antonucci1985}. This dusty torus, which is also the
main source of infrared emission in an AGN, can block the accretion
disk continuum and broad emission lines produced within the Broad Line
Region (BLR) by obscuring them with dusty gas
\citep[e.g.,][]{Sanders1989,Barvainis1990}.

A consequence of this geometry is that when the AGN is viewed face-on,
we observe a ``Type 1" active galaxy displaying both narrow lines
(permitted and forbidden) with line widths of hundreds of km s$^{-1}$,
as well as the broad lines emanating from the exposed hot ($\sim
10^{5}$ K) and bright photoionized gas around the central accretion
disk, with widths of thousands of km s$^{-1}$
\citep{Antonucci1993}. When viewed edge-on, the broad lines are
obscured by the optically thick dusty torus, and we observe only the
narrow lines generated on much larger scales characteristic of ``Type
2" active galaxies.

As a further clue to the structure of the torus, there is
observational evidence for a luminosity dependence of the fraction of
Type 1 and Type 2 AGNs. For example, in a deep hard X-ray survey,
\cite{Steffen2003} show that Type 1 AGNs dominate at higher
luminosities, while Type 2 AGNs become an important component of the
X-ray population only at lower, Seyfert-like luminosities. Similarly,
\cite{Hao2005b} shows that the fraction of Type 1 AGNs increases with
luminosity of the [OIII] narrow emission line (see also
\citealt{Simpson2005}), and studies using radio-selected AGN samples
consistently point to the same trend \citep[see e.g.,][]{Hill1996,
  Simpson2000, Grimes2004}.

Assuming the inner wall of the torus is set at the radius at which dust
sublimates \citep[e.g.,][]{Elvis1994}, then that distance is farther
out in luminous quasars (the more luminous relatives of Seyferts).  If 
the height of the torus for AGNs of varying luminosities remains
approximately constant, then the opening angle of the torus must
increase with luminosity. This leads to an expected dependence on
luminosity in the observed fraction of Type 1 and Type 2
AGNs. \cite{Lawrence1991} first suggested this idea, known as the
``receding torus" model (though see \citealt{Lawrence2010} for a
different perspective). It is not clear why the height of the torus
would be constant in all objects, but nevertheless a luminosity
dependence is expected, as it would be highly unlikely for the height
to vary in such a way as to produce a constant opening angle
\citep{Simpson2005}.

This highlights the need for a more fundamental understanding of the
torus. As the structure at the interface between the AGN and its host
galaxy, the torus plays a vital role in the make-up of a quasar. Yet
in the past, it has typically been modeled as a static structure
\citep[e.g.,][]{Pier1992,Fritz2006,Nenkova2008a,Schartmann2008,Honig2010}
which is likely only a rough approximation to the active inner region
surrounding an accreting black hole. Indeed, for a geometrically thick
torus to be supported hydrostatically, it must have a temperature on
the order of $10^{6}$ K or higher; dust in such a hot, high-density
environment would most certainly be destroyed
\citep{Dullemond2005}. However, dust signatures such as the
``infrared hump" are observed \citep{Sanders1989}, along with silicate
emission and absorption features
\citep[e.g.,][]{Siebenmorgen2005,Hao2005}. Therefore, dust near the
sublimation radius is certainly present in a wide variety of AGNs.
One way of getting around this inconsistency is to generate a static,
geometrically thick torus through infrared radiation pressure
\citep[e.g.,][]{Krolik2007,Shi2008}; however, this treatment neglects
the effect of radiation pressure from UV photons.

One way to help us understand the structure and physics of the dusty
torus is to examine the IR regime of quasar SEDs. Dust reprocesses a
significant fraction of the accretion power in the UV and X-ray, and
re-emits it in the infrared, acting as a sort of bolometer for the
quasar. Indeed, the \mbox{1 -- 100~$\um$} IR ``hump" of a quasar
spectrum accounts for nearly 40\% of the bolometric luminosity
\citep[e.g.,][]{Elvis1994,Richards2006}.  As commonly seen with {\em
Spitzer}, AGN IR spectra also show specific structures, such as
10~$\mu$m silicate features, that any successful model must explain.
While Type~2 objects commonly show silicate in absorption, emission is
often observed in Type~1 objects
\citep[e.g.][]{Siebenmorgen2005,Hao2005}.

High spatial-resolution imaging adds further requirements to any
successful torus model.  Using the \emph{Very Large Telescope
Interferometer} (VLTI), \cite{Jaffe2004} obtained interferometric
observations of NGC 1068 from 8.0 to 13.0 $\um$ that show different
dust temperatures found within close proximity; this hints at
structure within the torus \citep{Schartmann2005}.  Similarly, more
recent VLTI observations (from 8.0 to 13.0 $\um$) of the Circinus AGN
by \cite{Tristram2007} show irregular behavior (such as fluxes with
rapid angular variations) that can be explained by clumpiness.
Further, they find that the radial temperature dependence of the dust
indicates that some of the outer dust is exposed directly to radiation
from the nucleus. Together, these observations indicate that the dust
distribution in the torus is of clumpy or filamentary composition.

Previous researchers have tried different approaches to model these
observations of the torus; most of that work pictures the torus as
made up of clouds \citep[e.g.,][]{Nenkova2008a}, while some consider a
continuous structure \citep[e.g.,][]{Krolik2007}, although most of the
models are static.  In the quasi-clumpy models of
\cite{Dullemond2005}, ``clumps" represented by axisymmetric rings of
material are used to understand the differences between smooth and
clumpy models. Assumptions of their model include the idea that all
clumps have equal optical depth and a Gaussian density profile. They
found that several of their smooth and clumpy models were able to
suppress the 10~$\um$ silicate feature in emission. Thus they note
that although their results corroborate the idea that clumpy tori can
account for this observational feature, it does not unequivocally
confirm the idea of a clumpy torus, as families of continuous winds
models can explain it as well.

The formalism of \cite{Nenkova2008a} takes into account the recent
interferometric and spectroscopic results. Their model, composed of
$\sim 5-15$ dusty clouds along each radial equatorial ray, is
successful at explaining several puzzling aspects of AGN infrared
observations.  Specifically, dust on the bright side of one of their
optically thick clouds is much hotter than on the dark side, and dust
on the dark side of a cloud nearer the source of the illuminating
continuum can be as hot as dust on the bright side of one further
out. The clumpy model furthermore reproduces the behavior of the
10~$\um$ silicate feature, namely the lack of very deep absorption
features.

Clumpy-torus models are clearly useful; \cite{Schartmann2008} found
with their clumpy tori models that the existence of the 10~$\um$
silicate feature, whether in emission or absorption, depends on the
size, optical depth, and distribution of clouds closest to the
nucleus.  However, the family of cloud (or clumpy) torus models
described above typically neglects the dynamics of the clouds; a
notable exception to this has been work on accreting clouds \citep[see
e.g.,][]{Vollmer2004,Beckert2004,Schartmann2011}.  In addition, some
researchers have postulated that these clouds are entrained in a wind
\citep[e.g.,][]{Elitzur2006}.

\subsection{Dynamical Models of the Dusty ``Torus''}

The above-mentioned static models of the torus do not address the
question of why the torus has the structure it does, how it lofts gas
and dust up to large heights above the accretion disk, or why its
covering fraction may depend on luminosity.  To address these
questions, a robust, dynamical model must be developed that can
accurately reproduce the features that we observe in AGNs.  Such
dynamical models have been approached in a number of ways.  One
scenario, first proposed by \cite{Krolik1988}, suggests that the torus
consists of a number of optically thick clumps which orbit the central
engine and collide regularly with other clumps.

In a different dynamical paradigm, \cite{Dopita1998}, aiming to
describe the \emph{IRAS} colors of Seyfert galaxies, suggested that
the torus is a large-scale accretion flow of some continuous medium
towards the nucleus. This free-falling ``envelope" of material rotates
slowly and circularizes once it reaches the centrifugal radius. Since
their lower limits on the accretion rates are well above that which
would support the Eddington luminosity of the central engine, they
determined that much of the infalling material would flow away from
the accretion disk in a wind.  \citet{HopkinsEtAl2011} have recently
investigated this kind of dynamical model in numerical simulations,
examining larger-scale gas flows as a possible model for the torus.

Finally, another possibility is that the torus is a dusty wind flowing
from the accretion disk. Such an idea is supported by the growing
evidence for gaseous outflows in many types of AGNs.  Both radio-loud
and radio-quiet AGNs often exhibit blueshifted absorption lines that
can be broad and/or narrow; these features are evidence of outflowing
material.  Approximately 15\% of radio-quiet quasars display strong,
blueshifted absorption features at UV resonance transitions with
velocities as high as 0.1$c$ \citep[e.g.,][]{Reichard2003,
Gibson2009}.  There has also been observational evidence that suggests
the mass outflow rate in AGNs is nearly equal to the mass inflow rate
\citep[see e.g.,][]{Crenshaw2003, Chartas2003}.

\cite{Konigl1994} first proposed that the torus can be explained as a
disk-driven hydro-magnetic wind. They were motivated to consider such
winds due to their apparent success at explaining particular radiative
characteristics of young stellar objects (YSOs), which are similar in
some respects to those of AGNs. For example, they note that both types
of objects often exhibit flat IR spectra, strong Ca III triplet lines,
and broad Na D emission -- observational findings which have been
interpreted in terms of a strong central UV source surrounded by a
disk-like, dusty mass distribution. The hydro-magnetic wind model was
later expanded upon by \cite{Everett2005} to include a more realistic
treatment of radiative acceleration in magnetic winds.  

In addition, models where {\em infrared} radiation pushes a continuous
wind vertically off of the accretion disk beyond the dust sublimation
radius \citep[e.g.,][]{Krolik2007,Shi2008,Dorod2011a, Dorod2011b} have
also been examined.
% some changes by J. Everett, below 
This vertical, infrared radiation pressure may indeed be present, but
the radiation force due to the central, UV-bright AGN continuum has
notably more integrated power to drive dust grains than the IR
continuum; whereas, for instance, \citet{Dorod2011b} find a radiation
pressure on dust from infrared photons can increase the effective
Eddington ratio by a factor of approximately 10--30, our calculations
show that, when including the entire AGN continuum, that factor can
become $\sim 200$ \cite[this is somewhat less, but comparable to the
factor of $\sim 800$ enhancement found in the early models
of][]{Konigl1994}.
% end of changes

This paper will focus on the dynamical model of the torus as a dusty
wind, which is launched by both magnetohydrodynamical (MHD) forces as
well as radiation pressure due to the accretion disk continuum.  The
dusty wind generated in this manner can cover a large fraction of the
sky as seen from the central black hole. This model has the benefit of
being more self-consistent and including the important physics of
motion around the black hole and radiative acceleration. Ultimately,
we wish to use this model to understand how the physical properties of
dusty winds in AGNs correlate with their observable spectral
signatures in the IR.

% =====================================================
% STRUCTURE OF TORUS AND WIND MODEL
% =====================================================
\section{The Structure of the Torus and the Wind Model}

Our model of the torus consists of a dusty wind driven by both
hydro-magnetic forces and radiative acceleration. The model and its
corresponding code has been expanded upon from its original form
\citep[detailed by][]{Konigl1994} by \cite{Everett2005}, where a
comprehensive account of the model's components and key equations are
described. In this model, we advance on \cite{Everett2005} by adding
the continuum opacity of ISM dust grains, as specified by the ISM dust
model in \verb$Cloudy$ (version 06.02.09b, last described by
\citealt{Ferland1998}; for the dust model, see
\citealt{Mathis1977,vanHoof2004}).

As in previous magneto-centrifugal models \citep[see, e.g.,][hereafter
BP82]{Blandford1982}, the model assumes a parsec-scale magnetic field,
approximately vertical to the accretion disk. (It is important to
point out that the origin of such a field is not clear, although
simulations of jet launching from AGN accretion disks seem to favor
the advection of magnetic flux from large scales; this is a topic of
great interest in current research, see e.g., \citealt{Hawley2011}.)
With such a field, gas and dust particles orbiting in the accretion
disk are subject to the centrifugal force, pointing outwards along the
disk, as well as the gravitational force directed inwards towards the
central black hole. However, a charged particle, tied to the magnetic
field, will be flung outward along the magnetic field line if the
centrifugal acceleration overtakes the gravitational acceleration;
this can occur if the angle of the magnetic field line to the vertical
is $\gtrsim 30^{\circ}$ (BP82).
In this sense, the ``foundation'' of our model is a straightforward
application of the self-similar model of BP82 and \citet{Konigl1994}.
Such self-similar models allow relatively quick calculations of the
wind geometry and dynamics by intrinsically assuming that the shape of
the magnetic field lines (which change due to radiation pressure)
scales as radius, such that the wind geometry at large distances has a
similar shape to the wind geometry at small radii, only scaled up to a
larger size.  This self-similarity is a significant assumption, but it
allows for the calculation of both the radial and vertical momentum
equations, and in this case, allows for the simple addition of the
radiative acceleration by effectively decreasing the gravitational
potential \citep[see the Appendix in][]{Everett2005}.

As part of the assumption of self-similarity, the model requires that
the density at the surface of the accretion disk scales as a function
of radius, so that $\rho_0 \propto r^{-b}_0$ and $B_0 \propto
r^{-(b+1)/2}_0$, where the subscript `0' denotes values at the disk
surface.  Our particular implementation of the wind model allows for
any radial scaling parameter, $b$, but in all of the calculations
presented here, we set $b = 3/2$ (this is equivalent to assuming a
stationary accretion disk flow, where the mass-loss does not change
across different decades in disk radius).  Two other key parameters
that are set in this model are: $\lambda$, the ratio of total specific
angular momentum in the wind to that in the disk, and $\kappa$, the
dimensionless ratio of mass flux to magnetic flux; as in the
`standard' model of BP82, we set these parameters to $\lambda = 30$
and $\kappa = 0.03$.  These parameters help set the magnetic field
strength in the wind at the surface of the disk; as opposed to jet
solutions of BP82 which required $B \sim 10^2$ to $10^4$\,G, the
strongest field in our models was of order $0.03$\,G at the base of
the wind.

Once the wind is launched by centrifugal acceleration, it is also
subject to radiative acceleration from the photons emitted from the
accretion disk. The radiation pressure on the dust from the central
source is very strong -- even for cases where L/L$_{\rm Edd} = 0.1$,
it is approximately 10 times the force required to unbind dust from
the gravitational potential \citep{Everett2009}.
(Though line scattering is included in the calculation, the continuum
source dominates the line-radiation pressure in the dusty medium.)
Dust grains absorb
radially (anisotropically) streaming photons originating in the accretion
disk, and then de-excite isotropically. The force, felt by the dust particles
due to conservation of momentum, feeds back on the wind structure by bending
the magnetic wind radially away from the source. The radiative force therefore
works in conjunction with the magneto-centrifugal forces to accelerate the wind
flow off the accretion disk, and modifies the structure of the outflow.

The assumed dust composition in the model is a standard interstellar
distribution, with a mixture of silicates and graphites and a
continuous distribution of grain sizes. These dust grains will be
destroyed at temperatures greater than their sublimation temperature,
$T_{sub} \sim$ 1500 K. The wind's innermost radius for dust driving is
therefore set approximately to the dust sublimation radius, outside of
which the silicate and graphite dust can survive. The sublimation
radius is dependent on a number of factors, including the bolometric
luminosity and shape of the incident continuum.

\cite{Schartmann2008} note during their investigations of the dusty
torus as a clumpy medium that it is important to have a variety of
sublimation radii for different sizes of dust grains in order to
accurately model the IR SED from the dust. Their solution was to
utilize a model with three different grain species, each with five
different grain sizes.  Such a multi-component grain distribution
would be important, but given the complexity of our models and the
large amount of time required for computing each SED in even the
simplified single-grain case, we will use a single grain type and
sublimation radius as our first approximation.

% =====================================================
% MODEL CODE
% =====================================================
\subsection{The Model Code}\label{Code}

The model code works in two separate steps. First, the structure of the wind
is determined semi-analytically, starting with a pure MHD wind and then adding
in radiation pressure from dust and atomic lines. Second, a Monte Carlo
simulation uses the information about the structure of the wind to generate a
spectral energy distribution from $0.1 - 2000 \um$. As we are interested
primarily in the more readily observable IR regime, we plot these results from
$2 - 100 \um$. Furthermore, at wavelengths $<2 \um$, we cannot be certain of
the accuracy of our SED calculations, as the accuracy is affected by the
parameters chosen for the Monte Carlo simulation (see \textsection
~\ref{MC3Dtest}).

\begin{figure}[htp]
  \begin{center}
    \includegraphics[width=0.45\textwidth]{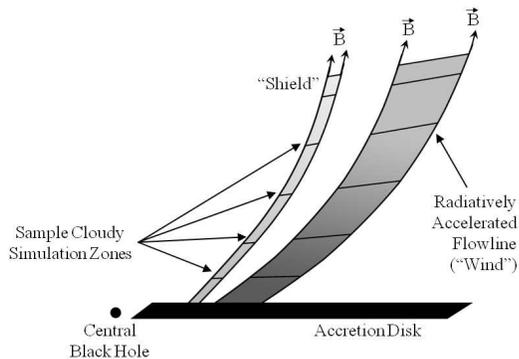}
  \end{center}
  \caption{Schematic diagram of the structure of the radiatively
    accelerated magneto-centrifugal wind model
    \citep{Everett2005}. The sub parsec-scale accretion disk is the
    source of the continuum illuminating the wind. The wind has two
    distinct components. The leftmost, thinner component is the
    ``shielding gas," and is modeled as a pure magneto-centrifugal
    wind. The wider streamline on the right is the dusty wind which is
    propelled by both magneto-centrifugal and radiative
    acceleration. The black lines within the shield and streamline are
    examples of the zones in which Cloudy determines opacities
    and simulates photo-ionization. In our calculations, the shielding
    gas is set to have an effectively negligible column density so
    that the wind is dominated by the radiatively accelerated
    component. \label{fig1}}
\end{figure}

In more detail, the program starts by calculating the pure
magneto-centrifugal wind dynamics (see Figure~\ref{fig1};
\citealt{Everett2005}) without including the effects of radiation
pressure.  Parameters such as the black hole mass, column density at
the base of the wind and starting location of the wind are specified
here. In this phase, an estimation of the density and velocity
structure of the wind as a function of height is calculated by a
relatively simple semi-analytic model of the magneto-centrifugal
acceleration of gas along a streamline (this is described in detail in
\citealt{Everett2005}). Next, \verb$Cloudy$ \citep{Ferland1998} is
used to calculate the photo-ionization of the gas and determine
opacities throughout the structure of the wind, given the central
continuum. With all of this information, the radiation pressure on the
wind is calculated. This is exactly as described in
\cite{Everett2005}, except for the addition of the opacity of dust,
which is supplied via \verb$Cloudy$'s ``ISM'' dust model. This
information is passed back into the first step, where the MHD wind
structure is modified accordingly. This process is iterated until
convergence, which typically takes seven iterations.

\begin{figure}[htp]
  \begin{center}
    \includegraphics[width=0.45\textwidth]{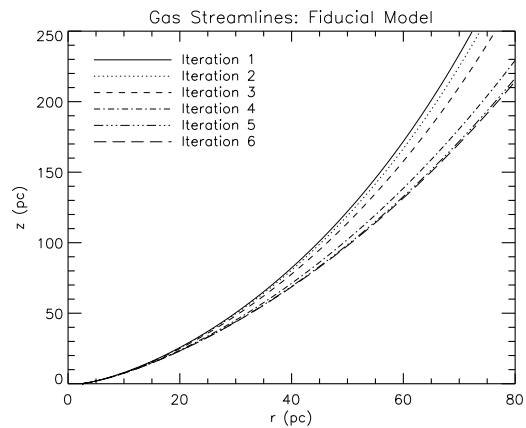}
  \end{center}
  \caption{Gas streamlines for the fiducial model with $N_{H,0} =
    10^{25}$ cm$^{-2}$, $L/L_{Edd} = 0.1$, $M_{BH} = 10^{8} \msun$,
    $\alpha_{ox} = 1.6$, and $i=60^{\circ}$. Streamlines are plotted
    in the $z$ direction (along the axis of symmetry) in units of
    parsecs, as a function of the $r$ (radial) direction, also in
    units of parsecs. The streamlines are shown for each iteration of
    the MHD radiative wind code. At each subsequent iteration (until
    convergence), the wind becomes more radial as radiative
    acceleration is taken into account. \label{streamlines}}
\end{figure}

Shown in Figure~\ref{streamlines} are the gas streamlines for our
fiducial model (with parameters as specified in \S~\ref{results}),
plotted in the $z$ direction along the axis of symmetry as a function
of the radial ($r$) direction. The streamlines are shown for each
iteration of the MHD radiative wind code. The radiation pressure acts
to bend the streamlines radially outward, away from the central
source, with each iteration, until the code converges.

\begin{figure}[htp]
  \begin{center}
    \includegraphics[width=0.45\textwidth]{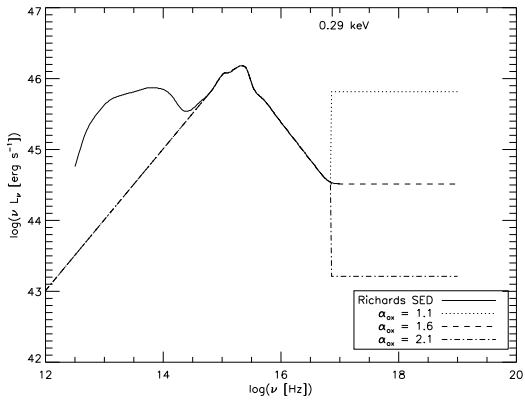}
  \end{center}
  \caption{The input continuum showing the range of $\alpha_{ox}$ values used.
    Luminosity is shown in units of log $\nu L_{\nu}$, and frequency is in
    units of log Hz. The solid line shows the original SED from
    \cite{Richards2006}, which includes the IR hump. The dotted, dashed and
    dashed-dotted lines display the modified continuum, with the IR hump
    replaced by a power law, and with an $\alpha_{ox}$ of 1.1, 1.6, and 2.1
    respectively. The discontinuity at $\nu = 7 \times 10^{16}$ Hz (or 0.29
    keV), where the X-ray continuum modification begins, is not physical, but
    serves to isolate the effect of the X-ray continuum from that of the
    far-UV.\label{aoxfig}}
\end{figure}

The second stage of our calculation uses the Monte Carlo simulation
program \verb$MC3D$ \citep{Wolf2003} to predict spectral energy
distributions. The input for the simulation is the output from the MHD
radiative wind code, which includes the structure of the wind and the
SED of the central continuum source (we use the composite SED from
\citealt{Richards2006}, modified to remove IR emission; see
Fig.~\ref{aoxfig} and \textsection ~\ref{CentralContinuum} for more
details). This input is passed to \verb$MC3D$: this is a general
purpose, 3D Monte Carlo code which calculates heating and emission
when given a central radiative source surrounded by a scattering and
absorbing medium.  \verb$MC3D$ first determines the temperature of the
dust grains due to heating from the continuum and reemission from
other dust grains. In a second step, \verb$MC3D$'s ray-tracer is then
used to create the SEDs from 0.1 -- 2000 $\um$, which we plot from 2
-- 100 $\um$.

% =====================================================
% METHODOLOGY
% =====================================================
\section{Methodology}\label{Method}

Initial tests for basic functionality, stability and consistency of the code
are documented in detail in \cite{Everett2005}. For this work, the code was
compiled on a number of different machines and it was verified that identical
results were produced on each machine. 
With confidence that the code performs as expected, we then investigated ways
to minimize computation time while maintaining an acceptable level of
accuracy. The bulk of the computing time is spent on generating the IR SED,
and thus we analyzed the \verb$MC3D$ input parameters to determine both
whether we could make our simulations more accurate without a significant
increase in time, and whether the process could be sped up significantly
without a loss of accuracy. These tests are reviewed in \textsection
~\ref{MC3Dtest}

After these basic considerations, we set out to explore the physical
parameter space that the model allows. AGNs exhibit a large range of
observed parameters, with varying luminosities, central continuum
shapes, central black hole masses, etc., and the way these parameters
interact in producing an IR SED is not necessarily simple or
obvious. Salient parameters were identified and tested against one
another in order to determine their effects on the dusty wind, and
therefore on the shape and power of the IR SED. An in-depth
description of each parameter examined can be found in \textsection
~\ref{ParamSpace} to \textsection ~\ref{results}.

\begin{figure}[htp]
  \begin{center}
    \includegraphics[width=0.45\textwidth]{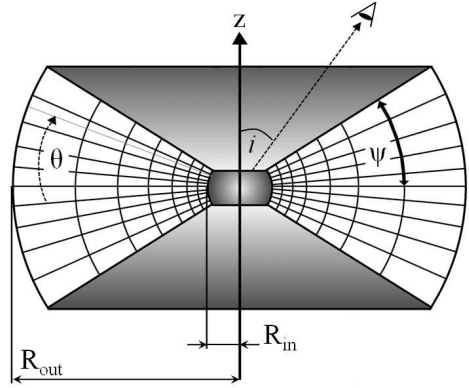}
  \end{center}
  \caption{MC3D model geometry. Listed in \cite{Wolf2003} as
    geometry \textbf{2D(b)}, it is a fully two-dimensional model, with
    vertical and radial dependences for the density distributions. The
    model is axisymmetric about the $z$ axis. Shown are the
    MC3D grid divisions in the $\theta$ and $r$ directions. The
    half-opening angle, $\psi$, is measured from the horizontal. The
    model radius is defined by $R_{out}$. The inclination angle, $i$,
    is defined from the normal, such that $i = 90^{\circ}$ is
    perpendicular to the axis of symmetry. \label{fig2}}
\end{figure}

\begin{figure}[htp]
  \begin{center}
    \includegraphics[width=0.45\textwidth]{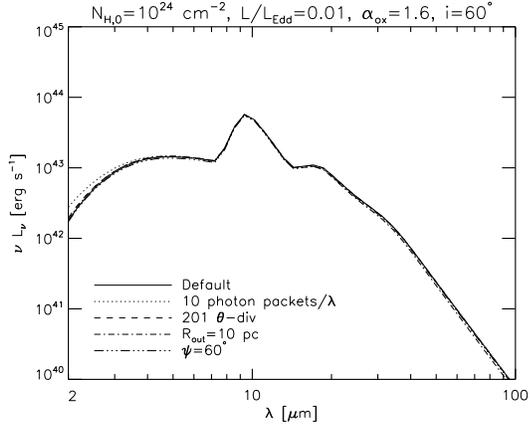}
  \end{center}
  \caption{Testing the effects of variation in the
    MC3D parameters. These IR SEDs are generated by MC3D with an
    inclination angle of $i = 60^{\circ}$, a base column density of
    $N_{H,0} = 10^{24}$ cm$^{-2}$, an Eddington ratio of $L/L_{Edd} =
    0.01$, a black hole mass of $M_{BH}$ = $10^{8} \msun$, and an
    $\alpha_{ox}$ value of 1.6. Luminosity is in units of $\nu
    L_{\nu}$, and wavelengths are in $\um$. Here, ``default" refers to
    the set of base testing parameters in \textsection~\ref{MC3Dtest},
    namely a model radius, R$_{out}$, of 20 parsecs with 30
    sub-divisions in the radial direction, 101 sub-divisions in
    the $\theta$ direction, a half opening angle, $\psi$, of 75
    degrees, and 100 photon packets per wavelength. Each subsequent
    SED varies one of these parameters, leaving the rest fixed at the
    default values. The only significant decrease in computation time
    arose from the change from 100 to 10 photons per wavelength, which
    is also the only result that deviated slightly from the SED
    generated by the base parameters. The largest discrepancy is seen
    at wavelengths $<3 \um$, where the accretion disk and host galaxy
    will also contribute to the observed emission. \label{mc3dparams}}
\end{figure}

% =====================================================
% MC3D TESTING
% =====================================================
\subsection{MC3D Testing Parameters}\label{MC3Dtest}

\verb$MC3D$ is a versatile code, with a range of customizable user-set
parameters for modelling dust-temperature distributions and SEDs. We
tested a number of these in order to determine an appropriate set of
base parameters which maintain a reasonable degree of accuracy while
minimizing the amount of time needed to generate the models.

\verb$MC3D$ has a number of model geometry options to choose from;
because our model is axisymmetric, we utilized the fully
two-dimensional model with a radial and vertical dependence for the
density distribution (see Figure~\ref{fig2}). Our default base values
for testing the effects of varying these parameters were as follows: a
maximum model radius, R$_{out}$, of 20 parsecs, with 30 sub-divisions
(grid points) in the radial direction and 101 sub-divisions in the
$\theta$ direction, a half-opening angle, $\psi$, of 75 degrees, and
100 photon packets per wavelength. We will discuss in more detail,
below, our tests of these parameter settings; for all of the parameter
changes discussed, Figure~\ref{mc3dparams} displays the differences in
the SED and Table~\ref{mc3d_table} details the effects on the emission
at 2, 3, 6, 10, and 50 $\um$ as compared to the default parameters.

For example, decreasing the number of photon packets per wavelength from the
default of 100 to 10 \citep{Wolf2003} had little effect on the SED, but
decreased the time required to compute the temperature distribution by almost
a factor of five. The variation that is seen between SEDs generated with 10 or
100 photon packets per wavelength is a minor loss of accuracy in the emission
from the hottest dust; the greatest difference is at $2 \um$, where the model
with 10 photon packets shows 1.61 times the emission as the model with 100
photon packets. As this loss is not relevant for our purposes -- the largest
discrepancy is seen at wavelengths $\lesssim$ 3 $\um$ where the accretion disk
and host galaxy typically contribute significantly to the observed emission -- and the increase
in computation efficiency is great, we opted to use 10 photon packets per
wavelength for all of our simulations.

Increasing the resolution of the grid by choosing 201 $\theta$ subdivisions
increased the computation time for the SED by several hours, with little
benefit. Again, the largest difference is seen at $2 \um$, where the test
model with 201 $\theta$ subdivisions showed 1.11 times the emission compared
to the default of 101 $\theta$ subdivisions.

Decreasing R$_{out}$ to 10 parsecs had little effect on the time for
generating the temperature distribution, but increased the time for generating
SEDs by more than a factor of 3. The difference in the SEDs is slight. The
greatest difference is seen at long wavelengths; at $50 \um$, the test model
with R$_{out}$ = 10 pc displays 0.88 times the emission compared to the
default model with R$_{out}$ = 20 pc.  We retain the value of 20 pc
for R$_{out}$.

Decreasing the half-opening angle from $\psi = 75^{\circ}$ to $\psi =
60^{\circ}$ also served to increase, rather than decrease, the computation
time. The SED showed a very slight overall change in normalization, with the
model with $\psi = 60^{\circ}$ showing a minor decrease in emission.

To summarize, for the rest of our models, we take the \verb$MC3D$ 
parameters as: 10 photon packets per
wavelength, 101 $\theta$ sub-divisions, R$_{out}$ = 20 pc, and $\psi =
75^{\circ}$.

A full list of these parameters, the range of values explored, and the
corresponding times to generate a temperature distribution and SED can be
found in Table~\ref{Table 1}. Most of the computations were done on a
workstation with a dual-core CPU, each at 3.16~GHz and a workstation with a
quad-core CPU, each at 2.83~GHz.

% =====================================================
% EXPLORING PARAMETER SPACE
% =====================================================
\subsection{Exploring the Parameter Space}\label{ParamSpace}

Turning to more physical parameters, we have investigated variations
in the final IR SED due to inclination angle of the observer ($i$),
the base column density of the wind ($N_{H,0}$), and changes to the
central continuum shape, luminosity, and black hole mass. All
parameter changes are specified before the wind structure is
calculated, with the exception of the inclination angle, which is only
specified when \verb$MC3D$ is called. We chose to run many of our
models at an inclination of $i=60^{\circ}$, as choosing inclination
angles smaller than that tended to increase the computation
time. Table~\ref{Table 2} details these parameters along with
corresponding relevant information, such as the terminal velocity of
the wind and the mass outflow rate, as calculated by the MHD wind
model, as well as the kinetic luminosity of the outflows.

\begin{figure}[htp]
  \begin{center}
    \includegraphics[width=0.45\textwidth]{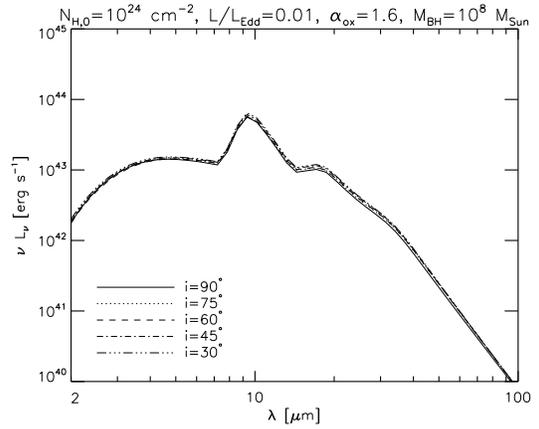}
  \end{center}
  \caption{Result of varying the observer's inclination angle, $i$,
    for a model with $N_{H,0} = 10^{24}$ cm$^{-2}$, $L/L_{Edd} =
    0.01$, $M_{BH}$ = 10$^{8} \msun$, and $\alpha_{ox} =
    1.6$. Luminosity is shown in units of $\nu L_{\nu}$, and
    wavelength in units of $\um$. The SEDs display remarkably little
    variation with changes in the observer's inclination
    angle. \label{n4e1x0_angle}}
\end{figure}

\begin{figure}[htp]
  \begin{center}
    \includegraphics[width=0.45\textwidth]{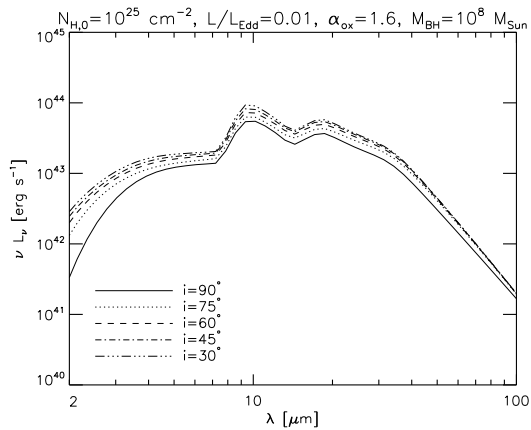}
  \end{center}
  \caption{Result of varying the observer's inclination angle, $i$,
    for a model with $N_{H,0} = 10^{25}$ cm$^{-2}$, $L/L_{Edd} =
    0.01$, $M_{BH}$ = 10$^{8} \msun$, and $\alpha_{ox} =
    1.6$. Luminosity is shown in units of $\nu L_{\nu}$, and
    wavelength in units of $\um$. At wavelengths $>5 \um$, the effect
    of decreasing the inclination angle is primarily a change in
    normalization, with smaller inclination angles displaying more
    power. At wavelengths $< 5 \um$, the effect is more pronounced,
    and the SEDs display a change in shape as well as normalization,
    with larger inclinations showing a steeper decrease in the amount
    of observable emission from the hottest
    dust. \label{n5e1x0_angle}}
\end{figure}

The relative power and shape of the SED were only slightly affected by
a change in the inclination angle, with larger inclination angles
showing a slightly lower normalization in the overall IR SED (see
Figures~\ref{n4e1x0_angle} and ~\ref{n5e1x0_angle}). This indicates
that the IR emission is consistent with being optically thin at most
wavelengths of interest.

The integrated luminosity varied significantly with changes in
the column density, with higher columns producing a higher overall
infrared luminosity. We chose to maintain a fixed column density
$N_{H,0} = 10^{25}$ cm$^{-2}$ for the remaining models, as this was
the only column which could match the order of magnitude of luminosity
comparable to a typical quasar ($L \approx 10^{45}$ erg
s$^{-1}$). Such a model generates mass outflow rates on order of $1
\msun$ yr$^{-1}$ (see Table~\ref{Table 2}).

% =====================================================
% CENTRAL CONTINUUM
% =====================================================
\subsection{The Central Continuum}\label{CentralContinuum}

\begin{figure}[htp]
  \begin{center}
    \includegraphics[width=0.45\textwidth]{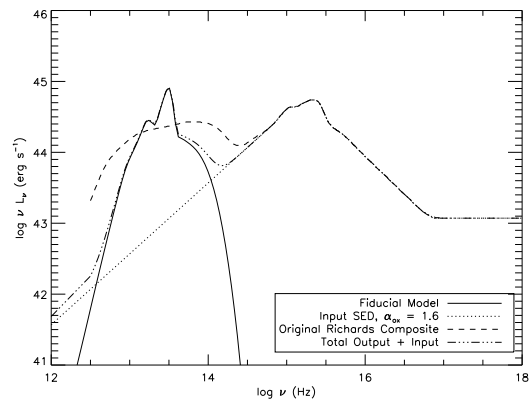}
  \end{center}
  \caption{Fiducial model with $N_{H,0} = 10^{25}$ cm$^{-2}$, $L/L_{Edd} =
    0.1$, $M_{BH}$ = 10$^{8} \msun$, $\alpha_{ox} = 1.6$, and $i=60^{\circ}$.
    Luminosity is shown in units of log $\nu L_{\nu}$, and wavelength in units
    of log Hz. The output SED for the fiducial model is denoted by the solid
    line. Also shown is the input SED (dotted line), the sum of the output and
    input SEDs (dashed-dot line), and the Richards composite SED (dashed
    line). \label{fiducial}}
\end{figure}

In order to investigate what sort of radiative output we would expect
to see from the dusty wind, we specify a central input continuum that
represents an approximation of the emission from the accretion disk
around the black hole.  We used a composite created from broadband
photometry of 259 optically luminous Sloan Digital Sky Survey (SDSS)
quasars, as described by \cite{Richards2006}, as a starting point for
this central continuum (see Figure~\ref{fiducial} for a plot of this
continuum, as well as the input continua modified from it).  The
optical and UV portions of the spectrum are assumed to be a fair
representation of the emission from the accretion disk. However, the
observed ``IR hump" (from 1 $\um$ -- 100 $\um$) in this composite
spectrum is due to the dust emission that we aim to model. Therefore,
the IR hump was removed and replaced with a simple power law
extrapolated from the optical continuum. Though the observed IR hump
from \mbox{1 -- 100~$\um$} is quite prominent, common features such as
the 10 and 18~$\um$ silicate emission features are not apparent, due
to a lack of spectral resolution in the broad-band photometry and the
effect of smearing out such features with composite
averaging. Nevertheless, such a composite SED serves as an excellent
basis for comparing to our IR SEDs models, as a check to ensure that
the power and general shape can be matched.  We will be testing
whether the wind model can account for the empirical IR hump.

% =====================================================
% CHANGES TO ALPHA OX
% =====================================================
\subsubsection{Changes to the Central Continuum: $\alpha_{ox}$}

The spectral index $\alpha_{ox}$ describes the amount of energy emitted in the
X-ray relative to the amount emitted in the UV/optical, and can be considered
a measure of the ``X-ray brightness" of a source. The parameter is defined as
follows \citep{Tananbaum1979}:

\begin{equation}
  \label{aox_eq}
  \alpha_{ox} = -0.384 \mathrm{\;log} \left[\frac{L_{\nu}(2~\mathrm{keV})}{L_{\nu}(2500~\mathrm{\AA})} \right]
\end{equation}

Quasars have been found with a range of $\alpha_{ox}$ values,
typically varying from 1.2 to 1.8. There is an anti-correlation
between $\alpha_{ox}$ and UV luminosity, as more luminous AGNs show
less X-ray emission relative to emission in the optical wavelengths
\citep[e.g.,][]{Just2007, Steffen2006}. In order to test this range of
$\alpha_{ox}$ values, we selected an ``average" value of 1.6
(appropriate for the UV luminosity of our fiducial model), and then
modified the input continuum (see Figure~\ref{aoxfig}) to reflect
extreme values of $\alpha_{ox} = 1.1$ (much higher than average output
in the X-ray), and $\alpha_{ox} = 2.1$ (much lower than average output
in the X-ray).  The X-ray modification begins at $7 \times 10^{16}$ Hz
and extends to $10^{19}$ Hz. The discontinuity at $\nu = 7 \times
10^{16}$ Hz is not physical, but serves to isolate the effect of the
X-ray continuum from that of the far-UV.

These new continua are passed as parameters to the wind and radiative
acceleration code. The results of these changes are described in
\textsection ~\ref{results}.

% =====================================================
% CHANGES TO L/LEDD
% =====================================================
\subsubsection{Changes to the Central Continuum: $L/L_{Edd}$}

Initial tests of the wind model were performed with an Eddington ratio of
$L/L_{Edd} = 0.01$. Such an Eddington ratio is characteristic of the Seyfert
regime, but may not accurately describe quasar luminosities. In order to test
the model at the higher luminosities more appropriate to quasars, we
incrementally increased the Eddington ratio to $L/L_{Edd} = 0.1$ by increasing
the total luminosity and keeping $M_{BH}$ fixed. Furthermore, we tested the
effects of changing black hole mass, which has the effect of changing the
Eddington luminosity, since:

\begin{equation}
  \label{Ledd_eq}
  L_{Edd} = 1.25 \times 10^{38} \left[ \frac{M}{\msun} \right] \mathrm{erg\;s}^{-1}
\end{equation}

We tested masses ranging from $M_{BH} = 10^{8} - 10^{9} \mathrm{\msun}$. We
note that although the part of the code that generates the wind structure was
able to accommodate such high luminosities with relative ease, the time for
generating the final SED increased by a factor of three or more at higher
luminosities.

% =====================================================
% RESULTS
% =====================================================
\section{Results}\label{results}

As a basis of comparison for many of our results, we adopt a fiducial model
with the following parameters: $N_{H,0} = 10^{25}$ cm$^{-2}$, $L/L_{Edd} =
0.1$, $M_{BH}$ = 10$^{8} \msun$, $i = 60^{\circ}$, and $\alpha_{ox} = 1.6$.
The Eddington luminosity for a black hole of mass $M_{BH}$ = 10$^{8} \msun$ is
$1.25 \times 10^{46}$ erg s$^{-1}$, giving a luminosity at 5100 \AA\ of
$L_{5100} = 3.7 \times 10^{29}$ erg s$^{-1}$ Hz$^{-1}$ for an Eddington ratio
of $L/L_{Edd} = 0.1$. Figure~\ref{fiducial} shows this fiducial model, plotted
with the input continuum, and the original \cite{Richards2006} composite SED.
The sum of the predicted IR SED and the input SED is also displayed. The composite
SED is normalized to the input SED at $L_{5100}$, and the output SED has been
integrated over $4 \pi$ steradians to match the units of the
observations.

With the base column density of $N_{H,0} = 10^{25}$ cm$^{-2}$, the
predictions approximately match the power expected from the
\cite{Richards2006} composite SED. The observed composite SED has
$\simeq 82 \%$ as much power radiated in the IR (from $2 - 100 \um$)
as in the optical/UV (from $2 \um - 1000$ \AA\.)  When normalized at
$L_{5100}$, our model with the smaller column density of $N_{H,0} =
10^{24}$ cm$^{-2}$, $L/L_{Edd} = 0.01$, $M_{BH}$ = 10$^{8} \msun$, $i
= 60^{\circ}$, and $\alpha_{ox} = 1.6$ has only 45\% as much power
radiated in the IR as in the composite optical/UV, which does not
match the expected output. Our model with a higher column of $N_{H,0}
= 10^{25}$ cm$^{-2}$ (and all other parameters as before) radiates
97\% as much power in the IR as in the composite optical/UV, which is
more than what is expected. Our fiducial model, with a higher column
of $N_{H,0} = 10^{25}$ cm$^{-2}$ and a higher Eddington ratio of
$L/L_{Edd} = 0.1$, has $\simeq 70 \%$ as much power radiated in the IR
as in the composite optical/UV, which is closer to the expected 82\%.

While the model SEDs can account for the gross power and shape of the
composite SED with the parameter values in the fiducial model, there
are specific features that do not match.  Specifically, the composite
has more power at both shorter (2--8~\micron) and longer ($>30$
\micron) wavelengths, and the fiducial model SEDs show more prominent
and sharply peaked silicate features at both 10 and 18 \micron.  On
this second point, note that the composite by construction smears out
spectral structure such as the silicate emission bumps as it is made
using broad-band photometry; many mid-IR spectra of luminous quasars,
however, show prominent silicate emission features \citep[see,
e.g.,][]{Siebenmorgen2005, Hao2007, Netzer2007, Mason2009,
Deo2011}. We also note that at long wavelengths, we do not expect much
of an AGN contribution from the dusty wind.  Empirically, far-infrared
emission of AGNs is typically weak unless the host is actively
star-forming \citep{Schweitzer2006,Netzer2007}.  In fact, the
composite SED contains little data beyond 24 $\um$, with that part of
the SED constructed by ``gap repair," replacing the missing data with
information extrapolated from \cite{Elvis1994} who used {\em IRAS}
photometry of low redshift quasars at these wavelengths.  We explore
these issues in more detail in \S~\ref{discussion}.

% =====================================================
% OUTFLOWS, VELOCITIES, LUMS
% =====================================================
\subsection{Mass outflows, terminal velocities, and kinetic luminosities}

Table~\ref{Table 2} contains information on the mass outflow rates
($dM/dt$), terminal velocities ($v_{term}$), and kinetic luminosities
($L_{kin}$) of each of the models. We see mass outflow rates ranging
from $0.96 - 4.19 \msun$ yr$^{-1}$, terminal velocities ranging from
$1904 - 7992$ km sec$^{-1}$, and kinetic luminosities ranging from
$1.13 \times 10^{42} - 7.71 \times 10^{43}$ erg s$^{-1}$. These values
are strongly affected by the black hole mass of the model: the largest
mass outflows, terminal velocities, and kinetic luminosities are
produced by the models with the highest black hole masses, even at
lower Eddington ratios.

% =====================================================
% VARYING i
% =====================================================
\subsection{Results of varying inclination angle}\label{inclination}

Here, we examine the result of varying the inclination angle from $i =
90^{\circ}$ (edge-on) to $i = 30^{\circ}$ (nearly
face-on). Inclination angles smaller than $i = 30^{\circ}$ were not
considered because of the great increase in time required to compute
the SEDs. As the first phase of our parameter exploration, these
models were run at luminosities $L/L_{Edd} = 0.01$ and $0.03$, lower
than that of our fiducial model.

Figure~\ref{n4e1x0_angle} shows the result of varying the inclination
angle for a model with $N_{H,0} = 10^{24}$ cm$^{-2}$, $L/L_{Edd} =
0.01$, $M_{BH}$ = 10$^{8} \msun$, and $\alpha_{ox} = 1.6$. This model
is different from our fiducial model in two ways: it has both a lower
column density ($N_{H,0} = 10^{24}$ cm$^{-2}$, as compared to $N_{H,0}
= 10^{25}$ cm$^{-2}$), and a lower Eddington ratio ($L/L_{Edd} =
0.01$, as compared to $L/L_{Edd} = 0.1$). Table~\ref{inclination_nh24}
details the amount of power emitted in the IR as compared to the
optical/UV for each model, as well as the ratios of emission at 3, 6,
10, and 50 $\um$, relative to the $i = 90^{\circ}$ model. Varying the
inclination angle for these parameters has almost no effect on the
shape or normalization of the SEDs. The lack of variation indicates
that IR emission must be optically thin at the wavelengths of
interest.

However, when we move to a higher base column density, variations in the SED
start to become apparent. Figure~\ref{n5e1x0_angle} shows the effects of
inclination angle for a model with $N_{H,0} = 10^{25}$ cm$^{-2}$, $L/L_{Edd} =
0.01$, $M_{BH}$ = 10$^{8} \msun$, and $\alpha_{ox} = 1.6$. This model is
different from our fiducial model as it has a lower Eddington ratio
($L/L_{Edd} = 0.01$, compared to $L/L_{Edd} = 0.1$). Variations in the SED are
much more apparent than they were at the lower column density of $N_{H,0} =
10^{24}$ cm$^{-2}$. Table~\ref{inclination_nh25} details the amount of power
emitted in the IR as compared to the optical/UV for each model, as well as the
ratios of emission at 3, 6, 10, and 50 $\um$, relative to the $i = 90^{\circ}$
model.

At wavelengths $>5 \um$, the effect of decreasing the inclination
angle is primarily a change in normalization, with smaller inclination
angles (more face-on lines of sight) displaying more power. At
wavelengths $< 5 \um$, the effect is more pronounced, and the SEDs
display a change in shape as well as normalization, with larger
inclinations (more edge-on lines of sight) showing a marked decrease
in the amount of emission from the hottest dust. At $3\um$, the $i =
30^{\circ}$ SED displays 2.4 times more emission than the $i =
90^{\circ}$ SED, and 1.48 times more emission at $50\um$. The results
of varying inclination angle are similar when looking at the model
with a slightly higher Eddington ratio, $L/L_{Edd} = 0.03$, which is
still smaller than that of our fiducial model.

The measurable variation at higher column densities indicates that the
wind is becoming more optically thick --- especially at the shortest
wavelengths --- as the column increases.

% =====================================================
% VARYING ALPHA OX
% =====================================================
\subsection{Results of varying $\alpha_{ox}$}

\begin{figure}[htp]
  \begin{center}
    \includegraphics[width=0.45\textwidth]{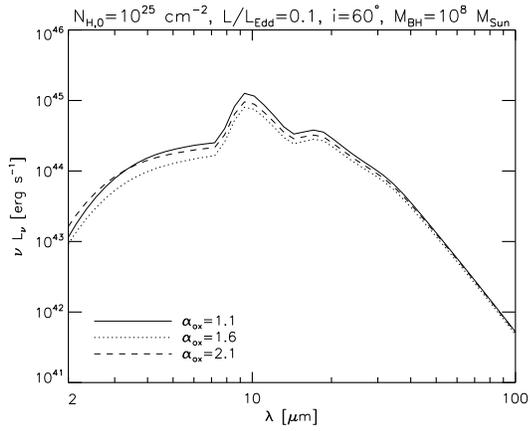}
  \end{center}
  \caption{Result of varying $\alpha_{ox}$ for a model with $N_{H,0} =
    10^{25}$ cm$^{-2}$, $L/L_{Edd} = 0.1$, $M_{BH}$ = 10$^{8} \msun$, and $i =
    60^{\circ}$. Luminosity is shown in units of $\nu L_{\nu}$, and wavelength
    is in units of $\um$. The model with $\alpha_{ox} = 1.1$ (solid line)
    displays the highest overall luminosity, with emission that decreases at
    wavelengths $<4 \um$. The model with $\alpha_{ox} = 1.6$ (dotted line)
    displays a slightly lower emission than both the X-ray brighter and X-ray
    fainter models. The model with $\alpha_{ox} = 2.1$ (dashed line) has a
    similar SED shape to the other models, but displays more emission at
    shorter wavelengths ($<4 \um$) than either of the other models. The shape
    and emission for wavelengths $> 50 \um$ is very similar for all
    models. \label{n5e4s6_aox}}
\end{figure}

Next, we examine the effects of changing the amount of energy emitted in the
X-ray relative to the amount emitted in the UV/optical, defined by parameter
$\alpha_{ox}$.
Figure~\ref{n5e4s6_aox} displays the results of varying $\alpha_{ox}$
as compared to our fiducial model, with $N_{H,0} = 10^{25}$ cm$^{-2}$,
$L/L_{Edd} = 0.1$, $M_{BH}$ = 10$^{8} \msun$, and $i =
60^{\circ}$. For reference, a $\Delta \alpha_{\rm ox}$ value of 1.0
corresponds to a factor of $\approx400$ difference in X-ray luminosity
at 2~keV for a set value of $L_{2500~\AA}$.  The depressed X-ray
emission of the X-ray faint model crudely mimics the effect of strong
X-ray absorption seen in some broad absorption line quasars
\citep[e.g.,][]{Gallagher2006}.  The X-ray bright model with
$\alpha_{ox} = 1.1$ displays the highest overall luminosity at
wavelengths $>4 \um$, with emission that decreases at wavelengths $<4
\um$, relative to the other models. Our fiducial model, with
$\alpha_{ox} = 1.6$, shows somewhat lower emission than both the X-ray
bright and X-ray faint models.

Table~\ref{aox_Ledd_0p1} shows the amount of power emitted in the IR
as compared to the optical/UV for each model, as well as the ratios of
emission at 3, 6, 10, and 50 $\um$, relative to the $\alpha_{ox} =
1.6$ fiducial model.  When the \cite{Richards2006} composite is
normalized to the $L_{5100}$ value of the input continuum, the
fiducial model displays 69.7\% as much power in the IR as compared to
the optical/UV power in the \cite{Richards2006} composite SED. The
$\alpha_{ox} = 1.1$ and $\alpha_{ox} = 2.1$ models display 102\% and
84.9\% as much power in the IR as compared to the composite
optical/UV, respectively.

The X-ray faint model with $\alpha_{ox} = 2.1$ displays more emission
at shorter wavelengths ($<4 \um$) than either of the other models, and
at wavelengths $>4 \um$ shows a change in normalization compared to
the other models. At wavelengths $> 50 \um$, the SEDs for all the
models appear to converge. Compared to the fiducial model with
$\alpha_{ox}=1.6$, the greatest difference is seen for the
$\alpha_{ox} = 1.1$ model which shows 1.55 times more emission at
$6\um$.

These results are perhaps counterintuitive, and reveal the complex
role of the SED in both accelerating and heating the dust.  The X-ray
bright model has significantly more power available to illuminate the
dusty medium, and therefore it makes sense that the IR SED is more
luminous relative to the other two models.  However, the relative lack
of emission at the shortest wavelengths in the X-ray bright model
indicates that there is a smaller volume of dust at the highest
temperatures, a natural consequence of a wind that is more radially
``combed out'' (and hence has a narrower wedge of material near the
sublimation temperature).  

In addition, the dust sublimation radius, $R_{\rm sub}$, is $\sim
13\%$ larger for the $\alpha_{\rm ox}$=1.1 models relative to the
$\alpha_{\rm ox}$=1.6 models (see Table~\ref{n5e4s6_aox}).  We
consider the 2\% difference in sublimation radii between the
$\alpha_{\rm ox}$=1.6 and 2.1 runs to be the same, within the
precision of the simulation.

These changes highlight the complex nature of how enhanced X-ray
emission impacts winds.  First, while X-rays contribute to heating,
their efficiency is quite low compared to UV photons: at 300 eV (near
the $\alpha_{\rm ox}$ jumps), the dust opacity (scattering plus
absorption) is about an order of magnitude below the UV opacity (see
Fig.~9 in \citealt{Draine2003}). Therefore, it requires a strong
increase in the X-ray flux to yield, e.g., a factor of two increase in
the near-IR flux.  A second effect is the increased momentum transfer
per photon of X-rays vs. UV light: X-rays are both more likely to
scatter (versus being absorbed) and carry more momentum.  Supporting
this interpretation, the X-ray-faint model with $\alpha_{\rm ox}=2.1$
shows the strongest short ($\lesssim4$~\micron) wavelength emission.
The UV power provides a ``floor'' to $R_{\rm sub}$; the smaller X-ray
power in the $\alpha_{\rm ox}=2.1$ model does not change the
sublimation radius significantly relative to the $\alpha_{\rm ox}=1.6$
model, while increasing the relative X-ray power by setting
$\alpha_{\rm ox}=1.1$ preferentially pushes out $R_{\rm sub}$ and
accelerates the wind more rapidly.  Both effects reduce the volume of
dust heated to the highest temperatures, and thus the relative power
coming out at the shortest IR wavelengths.

\begin{figure}[htp]
  \begin{center}
    \includegraphics[width=0.45\textwidth]{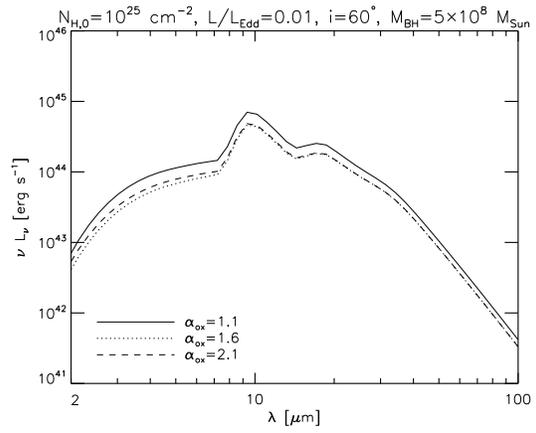}
  \end{center}
  \caption{Result of varying $\alpha_{ox}$ for a model with $N_{H,0} =
    10^{25}$ cm$^{-2}$, $L/L_{Edd} = 0.01$, $M_{BH}$ = $5\times10^{8} \msun$,
    and $i = 60^{\circ}$. Luminosity is shown in units of $\nu L_{\nu}$, and
    wavelength is in units of $\um$. The X-ray bright model with $\alpha_{ox}
    = 1.1$ (solid line) displays the highest overall luminosity. The model
    with $\alpha_{ox} = 1.6$ (dotted line) shows primarily a change in
    normalization, with little effect on the SED shape. The X-ray faint model
    with $\alpha_{ox} = 2.1$ (dashed line) also shows a change in
    normalization compared to the X-ray bright model, and is very similar to
    the model with $\alpha_{ox} = 1.6$, though it displays a slight increase
    in emission at wavelengths $<8 \um$. \label{n5e1m1s6_aox}}
\end{figure}

Figure~\ref{n5e1m1s6_aox} displays the results of varying
$\alpha_{ox}$ for a model with $N_{H,0} = 10^{25}$ cm$^{-2}$,
$L/L_{Edd} = 0.01$, $M_{BH}$ = $5\times10^{8} \msun$, and $i =
60^{\circ}$. This model has a lower Eddington ratio than our fiducial
model, as well as a higher black hole mass. The X-ray bright model
with $\alpha_{ox} = 1.1$ displays the highest overall luminosity.  The
model with $\alpha_{ox} = 1.6$ shows a change in normalization, with
less emission than the $\alpha_{ox}=1.1$ model, and little effect on
the SED shape.  Table~\ref{aox_Ledd_0p01} shows the amount of power
emitted in the IR as compared to the optical/UV for each model, as
well as the ratio of emission at 3, 6, 10, and 50 $\um$, relative to
the $\alpha_{ox} = 1.6$ model. When the \cite{Richards2006} composite
is normalized to the $L_{5100}$ value of the input continuum, the
model with $\alpha_{ox} = 1.6$ displays about 80\% as much power in
the IR as compared to the optical/UV power in the \cite{Richards2006}
composite SED. The $\alpha_{ox} = 1.1$ and $\alpha_{ox} = 2.1$ models
display $\sim120$\% and $\sim90$\% as much power in the IR as compared
to the composite optical/UV, respectively.

The X-ray faint model with $\alpha_{ox} = 2.1$ also shows a change in
normalization compared to the X-ray bright model, with less overall
emission, and is very similar to the model with $\alpha_{ox} = 1.6$,
though it displays a slight increase in emission at wavelengths $<8
\um$. Compared to the model with $\alpha_{ox}=1.6$, the $\alpha_{ox} =
1.1$ model shows 1.75 times more emission at $3\um$ and 1.27 times
more emission at $50\um$, and the $\alpha_{ox} = 2.1$ model shows only
1.20 times more emission at $3\um$ and displays the same amount of
emission at $50\um$.  As in the previous set of models, the impact of
changing $\alpha_{\rm ox}$ from 1.6 to 1.1 is to push out $R_{\rm sub}$
by $\sim18\%$; the difference between the two X-ray fainter models of
2\% is not considered significant.

The difference in the wind response to varying $\alpha_{ox}$ models
when the values of $M_{BH}$ and $L/L_{Edd}$ are also changed
illustrates the interplay between these parameters, and the challenges
in isolating the effects of the SED.  An increase in $M_{BH}$ of a
factor of 5 coupled with a decrease in $L/L_{Edd}$ of a factor of 10
means that the luminosity is a factor of 2 fainter than in the
previous model described in this section.  A lower luminosity coupled
with a larger black hole mass generates a wind with more vertical
streamlines, because the net radial component of the ejection
(centrifugal plus radiative minus gravitational) force will be
reduced.  For example, in the $L/L_{\rm Edd}=0.1$ X-ray bright model,
the wind streamlines at large radii have an angle from the disk of
$63^{\circ}$ vs. $74^{\circ}$ for the lower luminosity, X-ray bright
model.  The more vertical wind structure of the lower luminosity
models keeps the dusty wind closer to the illuminating source for
longer, but apparently decreases the observed emission from the
hottest dust.  We consider it likely that this is an artifact of the
observer's inclination angle (see Figure~\ref{n5e1x0_angle}); the more
vertical (because of lower radial acceleration) wind will be more
optically thick at shorter wavelengths. The view of the hottest dust
is therefore obscured for an observer viewing angle of 60$^\circ$.

% =====================================================
% VARYING MBH
% =====================================================
\subsection{Result of varying the black hole mass}

We examine the effects of changing the mass of the central black hole, with
values from $M_{BH}$ = $10^{8} - 10^{9} \msun$. Altering $M_{BH}$ also changes
the Eddington luminosity (see Equation~\ref{Ledd_eq}). For $M_{BH}$ = $10^{8}
\msun$, the Eddington luminosity is $L_{Edd} = 1.25 \times 10^{46}$ erg
s$^{-1}$, which gives a luminosity at 5100 \AA\ of $L_{5100} = 3.7 \times
10^{28}$ erg s$^{-1}$ Hz$^{-1}$ for $L/L_{Edd} = 0.01$.
The luminosities scale linearly with the black hole mass. In these tests, we
keep $L/L_{Edd} = 0.01$ fixed.

\begin{figure}[htp]
  \begin{center}
    \includegraphics[width=0.45\textwidth]{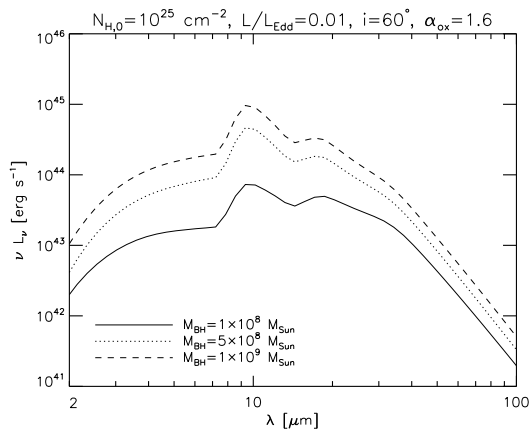}
  \end{center}
  \caption{Result of varying $M_{BH}$ for a model with $N_{H,0} = 10^{25}$
    cm$^{-2}$, $L/L_{Edd} = 0.01$, $\alpha_{ox}=1.6$, and $i = 60^{\circ}$.
    Luminosity is shown in units of $\nu L_{\nu}$, and wavelength is in units
    of $\um$. The highest black hole mass, $M_{BH}$ = $10^{9} \msun$ (dashed
    line), shows an increase in luminosity at all wavelengths compared to the
    model with $M_{BH}$ = $5\times10^{8} \msun$ (dotted line) and $M_{BH}$ =
    $10^{8} \msun$ (solid line). Here, higher black hole masses produce higher
    overall luminosities. The higher $M_{BH}$ models show a more peaked SED
    shape from $3 - 30 \um$ as compared to the $M_{BH}$ = $10^{8} \msun$
    model, as well as the overall increase in luminosity. \label{n5e1x0_mbh}}
\end{figure}

Figure~\ref{n5e1x0_mbh} shows the effect of changing the black hole
mass, which changes the luminosity for a fixed $L/L_{Edd}$. In this
situation, higher black hole masses produce overall more luminous
SEDs. For both $M_{BH}$ = $5\times10^{8} \msun$ and $M_{BH}$ = $10^{9}
\msun$, the shape and normalization of the SED is affected from $2 -
40 \um$ as compared to the $M_{BH}$ = $10^{8} \msun$
model. Table~\ref{mbh_table} shows the amount of power emitted in the
IR as compared to the optical/UV for each model, as well as the ratios
of emission at 3, 6, 10, and 50 $\um$, relative to $M_{BH}$ = 10$^{8}
\msun$. The higher $M_{BH}$ models show a more peaked SED shape from
$3 - 30 \um$ as compared to the $M_{BH}$ = $10^{8} \msun$ model, as
well as the overall increase in luminosity. Compared to the model with
$M_{BH}$ = $10^{8} \msun$, the model with $M_{BH}$ = $5\times10^{8}
\msun$ displays $\sim60$ times the emission at $10\um$ and $\sim80$
times the emission at $50 \um$, whereas the model with $M_{BH}$ =
$10^{9} \msun$ displays $\sim130$ times the emission at $10\um$, and
$\sim40$ times the emission at $50 \um$.  Given that the emission at
these comparison wavelengths is not just scaling linearly with the
luminosity (which is increasing by a factor of 5 and 10 as $M_{BH}$ is
increased by the same factors), the change in black hole mass clearly
has a direct influence on the IR power.

% =====================================================
% VARYING LEDD
% =====================================================
\subsection{Result of varying $L/L_{Edd}$}

We examine the effects of changing the Eddington ratio, $L/L_{Edd}$, which,
for a given black hole mass, has the effect of changing the bolometric
luminosity. For $M_{BH} = 10^{8} \msun$, an $L/L_{Edd}$ ratio of 0.1 gives a
luminosity at 5100 \AA\ of $L_{5100} = 3.7 \times 10^{29}$ erg s$^{-1}$
Hz$^{-1}$; $L/L_{Edd} = 0.07$ gives $L_{5100} = 2.6 \times 10^{29}$ erg
s$^{-1}$ Hz$^{-1}$; $L/L_{Edd} = 0.03$ gives $L_{5100} = 1.1 \times 10^{29}$
erg s$^{-1}$ Hz$^{-1}$; and an $L/L_{Edd} = 0.01$ gives $L_{5100} = 3.7 \times
10^{28}$ erg s$^{-1}$ Hz$^{-1}$.

\begin{figure}[htp]
  \begin{center}
   \includegraphics[width=0.45\textwidth]{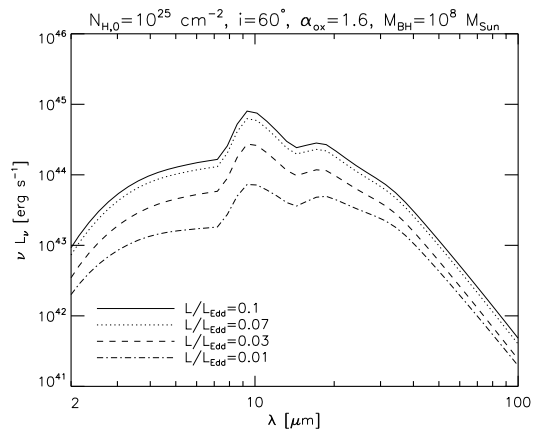}
  \end{center}
  \caption{Result of varying $L/L_{Edd}$ for a model with $N_{H,0} = 10^{25}$
    cm$^{-2}$, $M_{BH} = 10^{8} \msun$, $\alpha_{ox}=1.6$, and $i =
    60^{\circ}$. Luminosity is shown in units of $\nu L_{\nu}$, and wavelength
    is in units of $\um$. The difference between $L/L_{Edd} = 0.1$ (solid
    line) and $L/L_{Edd} = 0.07$ (dotted line) is a normalization factor, with
    little change in the SED shape. More pronounced changes to the SED shape,
    in addition to a change in normalization, can be seen for $L/L_{Edd} =
    0.03$ (dashed line) and $L/L_{Edd} = 0.01$ (dashed-dot line), with the
    latter displaying relatively more emission than the former at wavelengths
    $<6\um$. \label{n5x0_Ledd}}
\end{figure}

Figure~\ref{n5x0_Ledd} displays the effects of varying $L/L_{Edd}$ (changing
the luminosity). As is expected, higher $L/L_{Edd}$ ratios display a greater
luminosity as compared to lower $L/L_{Edd}$ ratios. The $L/L_{Edd} = 0.1$ and
$0.07$ models are very similar in shape, differing only by a normalization
factor. The $L/L_{Edd} = 0.03$ and $0.01$ models show a marked change in SED
shape as compared to the higher luminosity models, in addition to displaying a
lower overall luminosity. Table~\ref{Ledd_table} shows the amount of power
emitted in the IR as compared to the optical/UV for each model, as well as the
ratios of emission at 3, 6, 10, and 50 $\um$, relative to $L/L_{Edd} =0.1$. As
compared to the fiducial model with $L/L_{Edd} = 0.1$, the $L/L_{Edd} = 0.07$
model displays around 0.8 times the emission at each wavelength examined,
whereas the $L/L_{Edd} = 0.03$ model displays $\sim0.4$ times the emission at
$3\um$ and 0.5 times the emission at $50\um$.

\subsection{Distinguishing between $M_{BH}$ effects and $L/L_{Edd}$ effects}

\begin{figure}[htp]
  \begin{center}
    \includegraphics[width=0.45\textwidth]{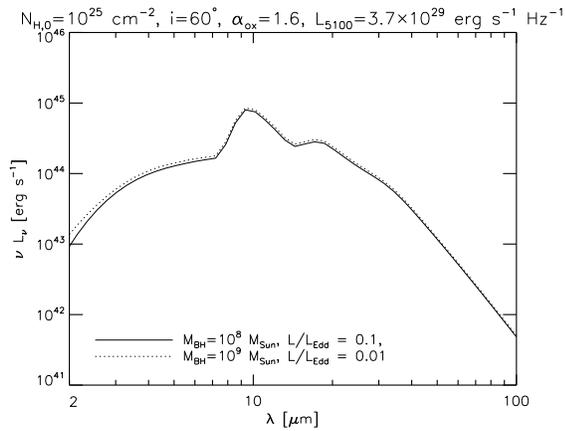}
  \end{center}
  \caption{Plot of models with the same input luminosity, $L_{5100} = 3.7
    \times 10^{29}$ erg s$^{-1}$ Hz$^{-1}$, for different Eddington ratios and
    black hole masses. The solid line shows our fiducial model, with $M_{BH} =
    10^{8} \msun$ and $L/L_{Edd} = 0.1$. The dotted line shows a model with
    $M_{BH} = 10^{9} \msun$ and $L/L_{Edd} = 0.01$. The two curves are not
    identical, with the model with $M_{BH} = 10^{9} \msun$ and $L/L_{Edd} =
    0.01$ showing relatively more power shortward of the
    10\micron\ silicate feature. 
     \label{distinguish}}
\end{figure}

As changing both $M_{BH}$ and $L/L_{Edd}$ effectively results in a
change in luminosity, one might ask whether the changes to these
parameters are redundant. To isolate these effects, we plot two models
with the same input luminosity, $L_{5100} = 3.7 \times 10^{29}$ erg
s$^{-1}$ Hz$^{-1}$ (see Figure~\ref{distinguish}). The two plots are
very similar, but not identical: the model with a higher black hole
mass and lower Eddington ratio displays a higher normalization from $2
- 40 \um$. These results suggest that the properties are not
redundant, though running further models for comparison would be of
great benefit.

% =====================================================
% CONCLUSIONS AND DISCUSSION
% =====================================================
\section{Conclusions and Discussion}
\label{discussion}

We have investigated the parameter space of our model of the torus as
a dusty magneto-centrifugal wind with radiative acceleration. We have
examined a number of interesting and important parameters and 
illustrated how they affect the IR SED of an AGN.  We have verified
that with input SEDs of bolometric luminosities $L \approx 10^{44} -
10^{46}$ erg s$^{-1}$ and a base column density of
$10^{25}$~cm$^{-1}$, the code produced reasonable IR SEDs with
approximately the right shape and luminosity ($L \approx 10^{43} -
10^{45}$ erg s$^{-1}$) as expected from the \cite{Richards2006}
composite of optically luminous SDSS quasars.  This is a promising
result for a relatively simple model given that we have not attempted
any fitting. A benefit to our model is that we are able to see
directly the effects of various physical parameters on the final IR
SED, which ultimately will allow us to understand the physical
properties of the torus itself. By determining which physical
parameters have an observable effect on the IR SEDs, and narrowing
down these parameters to ascertain which ones allow us to generally
reproduce the power expected, we have established a reasonable
starting point from which we can expand and further refine our model.

We summarize our results as follows:

\begin{enumerate}
\item Varying the inclination angle has little effect at base column densities
  $\leq 10^{24}$ cm$^{-2}$, but becomes more pronounced at base column
  densities $\geq 10^{25}$ cm$^{-2}$. This indicates that the emission is
  becoming slightly more optically thick as the column
  increases.  The most salient difference (see Fig.~8) is a deficit at
  the shortest wavelengths for the larger inclination angles,
  indicating that emission from the hottest dust, at the smallest
  radii, is being blocked.

\item The parameter $\alpha_{ox}$, characterizing the amount of energy emitted
  in the X-ray relative to the amount emitted in the optical/UV, has slight
  but measurable effects on the IR SEDs.  The X-ray bright model, with
  $\alpha_{ox} = 1.1$, displays the highest overall luminosity, with emission
  that decreases at wavelengths $< 4 \um$. Both the X-ray bright
  ($\alpha_{ox}
  = 1.1$) and X-ray faint ($\alpha_{ox} =2.1$) models display marginally
  more emission than the fiducial model with $\alpha_{ox} = 1.6$.

\item The short wavelength ($<5$~\micron) emission appears to be
  sensitive to the relative power of X-rays relative to UV in the SED,
  offering promise for using the properties of the IR SED to constrain
  the strength and shape of the high energy continuum.  This
  illustrates the utility of models where the IR-emitting medium can
  respond dynamically to the input SED; in static models the incident
  spectrum is quickly thermalized and information on its
  shape is lost.

\item Higher black hole masses produce more luminous IR SEDs, for a fixed
  value of $L/L_{Edd}$. The SEDs for $M_{BH} \geq 5 \times 10^{8} \msun$ are
  more peaked from $\sim 3 - 30 \um$ as compared to the SEDs for $M_{BH} < 5
  \times 10^{8} \msun$.

\item Higher Eddington ratios produce more luminous IR SEDs. The SEDs for
  $L/L_{Edd} \geq 0.03$ are more peaked from $\sim 3 - 30 \um$ as compared to
  SEDs for $L/L_{Edd} < 0.03$.

\item Although changes to both $L/L_{Edd}$ and $M_{BH}$ are effectively
  changes to the overall input luminosity, these effects may not be redundant,
  and appear to be slightly distinguishable from each other. The model with a
  higher black hole mass and lower Eddington ratio displays a higher
  normalization from $2 - 40 \um$.
\end{enumerate}

At present, although our model is producing approximately the right
amount of power, it does not adequately reproduce the amount of
expected emission in the hot dust, at wavelengths $\lesssim 8
\um$. This may be due in part to inclination angle effects. As
discussed in \textsection ~\ref{inclination}, although inclination
angle has little effect at lower column densities, it begins to show
an effect at column densities of $N_{H,0} = 10^{25}$ cm$^{-2}$, with
smaller inclination angles displaying more power, particularly at
wavelengths $< 5 \um$. It is possible that this effect is amplified
not only by an increase in column density, but by an increase in
luminosity as well. We chose an inclination angle of $i = 60^{\circ}$
for the majority of our models in order to increase our computational
efficiency, which allowed us to run more models in a shorter period of
time. However, such an inclination angle is not likely to represent
the average for Type~1 objects; a more characteristic choice would have
been between $i = 45^{\circ}$ and $i = 30^{\circ}$. The choice of $i =
60^{\circ}$ means that we may be obscuring the innermost parts of the
wind, and therefore missing the hottest dust, in many of our models.

Another possibility is that our assumption of an ISM dust composition
is not entirely correct. We assume a single silicate dust
sublimation radius, which neglects the hot graphite grains that can exist at smaller radii, and thus we are missing that emission. Further
investigations into the details of our dust composition and
sublimation radius are therefore required. Changing grain composition
can also affect the profile of the silicate features.

The above simulations with \verb$MC3D$ required long run-times. The
fiducial model takes $\sim$3--4 days on a 4-core desktop machine to
construct an SED for a single viewing/inclination angle. Further,
smaller sublimation radii and smaller inclination angles require much
longer run-times. For example, our fiducial model with a modified
$R_{\rm sub} = 1.5675 \times 10^{18}~{\rm cm}$ requires $\sim$2--3
months on an 8-core desktop machine. Smaller sublimation radii
translate to higher grid resolutions in the grid setup procedure for
\verb$MC3D$. To explore the sensitivity of our results to the effects
of a smaller sublimation radius, we found it necessary to port
\verb$MC3D$ to run on the many-core symmetric multi-processing (SMP)
compute nodes of SHARCNET\footnote{SHARCNET is a consortium of
Canadian academic institutions who share a network of high performance
computers.}. The fiducial model now takes $\sim 19$ hours when using
32 nodes; the highest resolution run mentioned above takes slightly
longer than a week.

\begin{figure}[htp]
  \begin{center}
   \includegraphics[width=0.45\textwidth]{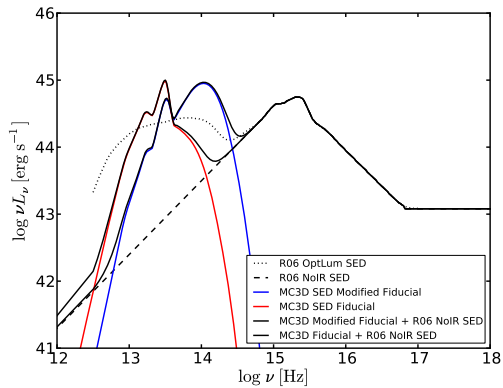}
  \end{center}
  \caption{The fiducial model with a smaller dust-sublimation radius shows an
    SED (blue solid line) peaking in the near-IR. As the dust-sublimation
    radius is increased the SED power decreases progressively in the near-IR
    and the peak emission shifts to longer wavelengths.\label{hot_dust_sed}}
\end{figure}

We modified our fiducial model, setting the inner radius of the dusty
wind to be $1.5675 \times 10^{18}~{\rm cm}$, about half that of the
original fiducial model, keeping all other parameters fixed. A maximum
temperature of 1615 K was reached within the dusty wind as modeled by
\verb$MC3D$, as opposed to approximately 800--1200 K in previous
variations of the fiducial model. This model reproduces the
blackbody-like shape of the near-IR emission between 1--8 \micron\
(see Figure~\ref{hot_dust_sed}) that is seen in mid-IR spectra of
luminous quasars \citep[e.g.,][]{Netzer2007,Deo2011}. The SED
corresponding to our fiducial run (red curve) and the high-resolution
run (blue curve), bracket the average near-IR emission in the
composite SED of Richards et~al. (2006). The composite mean SED is an
average over several important parameters such as inclination,
luminosity and $M_{BH}$, and for a more accurate comparison, one would
have to average over the similar range of models. That full study is
beyond the scope of this paper, but we find that, in our predicted
SEDs, the near-IR hot-dust luminosity decreases as we increase the
inclination angle, giving approximately the correct trend. These
results show that changes in the dust sublimation radius approximate
the expected presence of hot graphite grains, and corresponding grid
resolution (an effect of the choice of grid construction process) are
important effects to consider when comparing model SEDs to
observations.

We have explored a wider parameter space of this model based
originally on the hydro-magnetic wind model of \cite{Konigl1994}. Our
results are roughly consistent with theirs, with our models displaying
similar $10 \um$ micron emission to their model (labeled as `A2' in
\citealt{Konigl1994}), with $M_{BH} = 10^{7} \msun$, a bolometric
luminosity $L_{bol} = 10^{45}$ erg s$^{-1}$, and a mass outflow rate
of 1.0 $\msun$ yr$^{-1}$. We do not, however, see the silicate feature
in absorption, as they do with their model `A3', with $M_{BH} = 10^{7}
\msun$, a bolometric luminosity $L_{bol} = 10^{44}$ erg s$^{-1}$, and
a mass outflow rate of 3.0 $\msun$ yr$^{-1}$, even in our models with
extremely high mass outflow rates ($>4.0 \msun$ yr$^{-1}$) and overall
higher mass. \cite{Nenkova2008b} also see the silicate feature in
absorption with many of their clumpy tori models. This may in part
mean that a higher resolution is required to resolve the wind,
although models with a significantly increased resolution would also
require an increase in computation time (see above
discussion). Furthermore, \cite{Konigl1994} add in radiative
acceleration only approximately; it is possible that with more
realistic radiative acceleration added in, the wind accelerates off
the disk quickly enough that the density drops very quickly with
height, and only a very small region at the disk surface remains where
we might observe silicate absorption.

Curiously, we find that the shape of our SEDs, in general, are similar to
\cite{Schartmann2008}'s clumpy, face-on models, and are rather dissimilar to
their continuous models (see the upper left panel in Figure 10
of \citealt{Schartmann2008}). We assume that they take the same parameters as
described in \cite{Schartmann2005}: a black hole mass of $M_{BH} = 6.6 \times
10^{7} \msun$, and an $L_{disk}/L_{Edd}$ value of 0.06, parameters which are
descriptive of the lower luminosity Seyfert regime of AGNs. Although our
models are geared more towards describing luminous quasars, their model is
most comparable to our model with $M_{BH} = 1 \times 10^{8} \msun$ and
$L/L_{Edd} = 0.07$. Again, however, we do not see the silicate feature in
absorption as they do for edge-on models.

% =====================================================
% FUTURE WORK
% =====================================================
\subsection{Future Work}

Future work on this model will entail refining our parameters, such as
testing smaller inclination angles to determine whether we can more
accurately reproduce the emission from the hottest dust.  Along these
lines, we aim to decrease the computation time further, to enable us
to resolve the inner graphite-dominated region that likely gives rise
to the near-IR emission. Then, we will be able to explore the
importance of multiple dust components, at the necessary higher
resolution.  Further next steps will involve detailed comparisons of
the predicted SEDs with a broad sample of IR observations, for example
from the \textit{Spitzer Space Telescope}. Such in-depth comparisons
will allow us to further constrain the salient parameters that we have
isolated in this paper.

Currently, we are investigating the effects of increasing the
``shielding gas'' column density, $N_{sg}$. The high X-ray flux from
the AGN is prone to over-ionizing the gas in the wind. A thick layer
of hot, highly ionized ``shielding gas," located interior to the
wind-launching dust sublimation radius, is required in order to
prevent this over-ionization. Such a construct was first hypothesized
in the context of continuous quasar winds as the ``hitchhiking gas"
described by \cite{MCGV1995}, and later studied by \cite{Proga2004}
and \cite{Everett2005}. Subsequent empirical evidence for the
shielding gas was found in the signatures of high column density X-ray
absorption in broad absorption line quasars
\citep[e.g.][]{Gallagher2006}.  Modifications to the \verb$MC3D$ code
are required in order to deal with absorbed continua passing through
different column densitities of shielding gas to determine the effect
on the dust sublimation radius and output IR SED.

Finally, another step towards perhaps more realistic models would be
to simulate dusty winds of a filamentary nature.  Such a model may
better represent the structures seen in interferometric observations
of AGNS \citep[see, e.g.,][]{Tristram2007}.

% =====================================================
% ACKNOWLEDGMENTS
% =====================================================
\acknowledgments

We thank Dean Hines and Gordon Richards for their helpful feedback on
the manuscript before submission, and the comments of the anonymous
referee improved the presentation in the article.  The authors
acknowledge the financial support of the National Sciences and
Engineering Research Council of Canada (S.K.K., S.C.G., and R.P.D.)
and the Ontario Early ResearcherAward Program (S.C.G. and R.P.D.).
S.K.K. also gratefully acknowledges the financial support of the
Ontario Graduate Scholarship program.  J.E.'s work was supported by
grants NSF AST-0507367, NSF AST-0907837, and NSF PHY-0215581 and NSF
PHY-0821899 (to the Center for Magnetic Self-Organization in
Laboratory and Astrophysical Plasmas).  This research has made use of
NASA's Astrophysics Data System.  Support for the early stages of this
work, part of the {\em Spitzer} Space Telescope Theoretical Research
Program, was provided by NASA through a contract issued by the Jet
Propulsion Laboratory, California Institute of Technology under a
contract with NASA.  This work was made possible by the facilities of
the Shared Hierarchical Academic Research Computing Network
(SHARCNET:www.sharcnet.ca) and Compute/Calcul Canada.

\begin{deluxetable}{llllll}
  \tablewidth{0pt}
  \tablecaption{Variations in testing parameters for MC3D\label{mc3d_table}}
  % \tablecolumns{9}
  \tablehead{
    \colhead{Parameter Change} &
    \colhead{$2 \um$ Ratio} &
    \colhead{$3 \um$ Ratio} &
    \colhead{$6 \um$ Ratio} & 
    \colhead{$10 \um$ Ratio} &
    \colhead{$50 \um$ Ratio} 
  }
  \startdata
  Default                & 1.00 & 1.00 & 1.00 & 1.00 & 1.00 \\
  10 photons/$\lambda$   & 1.61 & 1.14 & 0.99 & 0.99 & 0.98 \\
  201 $\theta$ divisions & 1.11 & 1.04 & 1.00 & 0.99 & 1.00 \\
  $R_{out} = 10$ pc      & 1.04 & 1.04 & 1.02 & 0.99 & 0.88 \\
  $\psi = 60^{\circ}$    & 0.98 & 0.96 & 0.96 & 0.96 & 0.96 \\
  \enddata
  \tablecomments{All models are compared to the model with default base
    parameters of 100 photon packets per wavelength, 101 $\theta$
    sub-divisions, R$_{out}$ = 20 pc, and $\psi = 75^{\circ}$.}
\end{deluxetable}

\begin{deluxetable}{lllllllll}
  \tablewidth{0pt}
  \tablecolumns{9}
  \tablecaption{Testing Parameters\label{Table 1}}
  \tablehead{
    \colhead{Model} & 
    \colhead{Photon Packets} &
    \colhead{Model Radius} &
    \colhead{\# of $\theta$} &
    \colhead{$\psi$ \tablenotemark{1}} &
    \colhead{$i$\tablenotemark{2}} &
    \colhead{$N_{H,0}$\tablenotemark{3}} &
    \colhead{$t_{tem}$\tablenotemark{4}} &
    \colhead{$t_{SED}$\tablenotemark{5}} \\
    \colhead{Name} &
    \colhead{per $\lambda$} &
    \colhead{(pc)} &
    \colhead{Divisions} &
    \colhead{($^{\circ}$)} &
    \colhead{($^{\circ}$)} &
    \colhead{(cm$^{-2}$)} &
    \colhead{(hrs)} &
    \colhead{(hrs)} \\
  }
  \startdata
  n4testS1 & 100 & 20 & 101 & 75 & 60 & $10^{24}$ & 38 & 59  \\
  n4testS2 & 10  & 20 & 101 & 75 & 60 & $10^{24}$ & 8  & 58  \\
  n4testS3 & 100 & 20 & 201 & 75 & 60 & $10^{24}$ & 39 & 64  \\
  n4testS4 & 100 & 10 & 101 & 75 & 60 & $10^{24}$ & 38 & 187 \\
  n4testS5 & 100 & 20 & 101 & 60 & 60 & $10^{24}$ & 49 & 77  \\ 
  n5testS1 & 10  & 20 & 101 & 75 & 30 & $10^{25}$ & 4  & 31  \\ 
  n5testS2 & 10  & 20 & 101 & 75 & 45 & $10^{25}$ & 4  & 37  \\
  n5testS3 & 10  & 20 & 101 & 75 & 60 & $10^{25}$ & 4  & 34  \\
  n5testS4 & 10  & 20 & 101 & 75 & 75 & $10^{25}$ & 4  & 31  \\
  n5testS5 & 10  & 20 & 101 & 75 & 90 & $10^{25}$ & 4  & 44  \\
  \enddata
  \tablenotetext{1}{Half-opening angle}
  \tablenotetext{2}{Inclination angle}
  \tablenotetext{3}{Base column density}
  \tablenotetext{4}{Time to generate temperature distribution}
  \tablenotetext{5}{Time to generate SED}
\end{deluxetable}

\begin{deluxetable}{lllllll}
  \tablewidth{0pt}
  \tablecaption{Parameter Space\label{Table 2}}
  \tablehead{
    \colhead{Model} &
    \colhead{$M_{BH}$\tablenotemark{1}} &
    \colhead{$L/L_{Edd}$\tablenotemark{2}} &
    \colhead{$\alpha_{ox}$\tablenotemark{3}} &
    \colhead{$dM/dt$\tablenotemark{4}} &
    \colhead{$v_{term}$\tablenotemark{5}} &
    \colhead{$L_{kin}$\tablenotemark{6}} \\
    \colhead{Name} & 
    \colhead{($\times 10^{8} \msun$)} &
    \colhead{ } & 
    \colhead{ } &
    \colhead{($\msun$ yr$^{-1}$)} & 
    \colhead{(km sec$^{-1}$)} &
    \colhead{($\times 10^{42}$ erg s$^{-1}$)}
  }
  \startdata
  n5e1x0s6  & $1$               & 0.01 & 1.6 & 0.99      & 1904      & $1.13$                                    \\
  n5e1xas6  & $1$               & 0.01 & 1.1 & 0.96      & 2344      & $1.67$                                    \\
  n5e1xbs6  & $1$               & 0.01 & 2.1 & 0.99      & 1959      & $1.19$                                    \\
  n5e2x0s6  & $1$               & 0.03 & 1.6 & 1.00      & 2755      & $2.40$                                    \\
  n5e2xas6  & $1$               & 0.03 & 1.1 & 1.05      & 2884      & $2.76$                                    \\
  n5e2xbs6  & $1$               & 0.03 & 2.1 & 1.00      & 2754      & $2.39$                                    \\
  n5e3x0s6  & $1$               & 0.07 & 1.6 & 1.14      & 2877      & $2.97$                                    \\
  n5e4x0s6  & $1$               & 0.1  & 1.6 & 1.23      & 2899      & $3.25$                                    \\
  n5e4xas6  & $1$               & 0.1  & 1.1 & 1.32      & 3162      & $4.14$                                    \\
  n5e4xbs6  & $1$               & 0.1  & 2.1 & 1.22      & 2907      & $3.24$                                    \\
  % s2e4x0s6 & $10^{8}$          & 0.1  & 1.6 & $10^{22}$ & $10^{25}$ & 1.164   & 2556.2 & $2.397 \times 10^{42}$ \\
  % s2e4xas6 & $10^{8}$          & 0.1  & 1.1 & $10^{22}$ & $10^{25}$ & 1.315   & 3161.7 & $4.142 \times 10^{42}$ \\
  % s2e4xbs6 & $10^{8}$          & 0.1  & 2.1 & $10^{22}$ & $10^{25}$ & 1.211   & 2913.8 & $3.240 \times 10^{42}$ \\
  % s3e4x0s6 & $10^{8}$          & 0.1  & 1.6 & $10^{23}$ & $10^{25}$ & 1.192   & 2948.7 & $3.266 \times 10^{42}$ \\
  % s3e4xas6 & $10^{8}$          & 0.1  & 1.1 & $10^{23}$ & $10^{25}$ & 1.334   & 3129.9 & $4.118 \times 10^{42}$ \\
  % s3e4xbs6 & $10^{8}$          & 0.1  & 2.1 & $10^{23}$ & $10^{25}$ & 1.239   & 2870.6 & $3.217 \times 10^{42}$ \\
  % s4e4x0s6 & $10^{8}$          & 0.1  & 1.6 & $10^{24}$ & $10^{25}$ & 1.192   & 2948.5 & $3.266 \times 10^{42}$ \\
  % s4e4xas6 & $10^{8}$          & 0.1  & 1.1 & $10^{24}$ & $10^{25}$ & \nodata & \nodata                         \\
  % s4e4xbs6 & $10^{8}$          & 0.1  & 2.1 & $10^{24}$ & $10^{25}$ & \nodata & \nodata                         \\
  n5e1m1s6  & $5$               & 0.01 & 1.6 & 2.39      & 5994      & $27.1$                                    \\
  e1m1xas6  & $5$               & 0.01 & 1.1 & 2.51      & 5958      & $28.1$                                    \\
  e1m1xbs6  & $5$               & 0.01 & 2.1 & 2.38      & 6000      & $27.0$                                    \\
  n5e4m1s6  & $5$               & 0.1  & 1.6 & 3.80      & 4999      & $29.9$                                    \\
  e4m1xas6  & $5$               & 0.1  & 1.1 & 4.19      & 5268      & $36.6$                                    \\
  e4m1xbs6  & $5$               & 0.1  & 2.1 & 3.78      & 5006      & $29.9$                                    \\
  % s4e4m1s6 & $5 \times 10^{8}$ & 0.1  & 1.6 & $10^{24}$ & $10^{25}$ & 2.388   & 5916.9 & $2.635 \times 10^{43}$ \\
  n5e1m2s6  & $10$              & 0.01 & 1.6 & 3.82      & 7992      & $76.9$                                    \\
  e1m2xas6  & $10$              & 0.01 & 1.1 & 4.07      & 7758      & $77.1$                                    \\
  e1m2xbs6  & $10$              & 0.01 & 2.1 & 3.82      & 7991      & $76.8$                                    \\
  % s4e4m2s6 & $10^{9}$          & 0.1  & 1.6 & $10^{24}$ & $10^{25}$ & \nodata & \nodata                         \\
  \enddata
  \tablecomments{All models have a base column density of $N_{H,0} = 10^{25}$ cm$^{-2}$.}
  \tablenotetext{1}{Black hole mass}
  \tablenotetext{2}{Eddington ratio}
  \tablenotetext{3}{Amount of power in optical relative to X-ray}
  \tablenotetext{4}{Mass outflow rate}
  \tablenotetext{5}{Terminal velocity of outflow}
  \tablenotetext{6}{Kinetic luminosity of outflow}
\end{deluxetable}

\begin{deluxetable}{llllll}
  \tablewidth{0pt}
  \tablecaption{Variations in $i$ for $N_{H,0} = 10^{24}$ cm$^{-2}$\label{inclination_nh24}}
  % \tablecolumns{9}
  \tablehead{
    \colhead{Parameter Change} &
    \colhead{$L_{IR}/L_{opt/UV}$} & 
    \colhead{$3 \um$ Ratio} &
    \colhead{$6 \um$ Ratio} & 
    \colhead{$10 \um$ Ratio} &
    \colhead{$50 \um$ Ratio} 
  }
  \startdata
  $i = 90^{\circ}$ & 44.3\% & 1.00 & 1.00 & 1.00 & 1.00 \\
  $i = 75^{\circ}$ & 46.0\% & 1.03 & 1.04 & 1.04 & 1.11 \\
  $i = 60^{\circ}$ & 45.5\% & 0.96 & 1.04 & 1.02 & 1.12 \\
  $i = 45^{\circ}$ & 47.9\% & 1.02 & 1.07 & 1.09 & 1.14 \\
  $i = 30^{\circ}$ & 49.8\% & 1.07 & 1.09 & 1.15 & 1.15 \\
  \enddata
  \tablecomments{All models are compared to the model with $i=90^{\circ}$,
    $N_{H,0} = 10^{24}$ cm$^{-2}$, $L/L_{Edd} = 0.01$, $M_{BH}$ = 10$^{8}
    \msun$, and $\alpha_{ox} = 1.6$. The ratio $L_{IR}/L_{opt/UV}$ is the
    amount of power emitted in the IR from $2 - 100 \um$ by a certain model as
    compared to the amount of power emitted in the optical/UV (from $2 \um -
    1000$ \AA) by the composite continuum by \cite{Richards2006}, when both
    are normalized at $L_{5100}$.}
\end{deluxetable}

\begin{deluxetable}{llllll}
  \tablewidth{0pt}
  \tablecaption{Variations in $i$ for $N_{H,0} = 10^{25}$ cm$^{-2}$\label{inclination_nh25}}
  % \tablecolumns{9}
  \tablehead{
    \colhead{Parameter Change} &  \colhead{$L_{IR}/L_{opt/UV}$} & 
    \colhead{$3 \um$ Ratio} &  \colhead{$6 \um$ Ratio} & 
    \colhead{$10 \um$ Ratio} &  \colhead{$50 \um$ Ratio} 
  }
  \startdata
  $i = 90^{\circ}$ & 70.9\%  & 1.00 & 1.00 & 1.00 & 1.00 \\
  $i = 75^{\circ}$ & 85.3\%  & 1.59 & 1.14 & 1.14 & 1.34 \\
  $i = 60^{\circ}$ & 97.5\%  & 1.91 & 1.30 & 1.31 & 1.42 \\
  $i = 45^{\circ}$ & 106.6\% & 2.13 & 1.39 & 1.47 & 1.46 \\
  $i = 30^{\circ}$ & 114.7\% & 2.40 & 1.48 & 1.65 & 1.48 \\
  \enddata
  \tablecomments{All models are compared to the model with $i=90^{\circ}$,
    $N_{H,0} = 10^{25}$ cm$^{-2}$, $L/L_{Edd} = 0.01$, $M_{BH}$ = 10$^{8}
    \msun$, and $\alpha_{ox} = 1.6$. The ratio $L_{IR}/L_{opt/UV}$ is the
    amount of power emitted in the IR from $2 - 100 \um$ by a certain model as
    compared to the amount of power emitted in the optical/UV (from $2 \um -
    1000$ \AA) by the composite continuum by \cite{Richards2006}, when both
    are normalized at $L_{5100}$.}
\end{deluxetable}

\begin{deluxetable}{lrccccc}
  \tablewidth{0pt}
  \tablecaption{Variations in $\alpha_{ox}$ for $L/L_{Edd} = 0.1$ and $M_{BH}$ = 10$^{8} \msun$ \label{aox_Ledd_0p1}}
  % \tablecolumns{9}
  \tablehead{
    \colhead{Parameter Change} &
    \colhead{$L_{IR}/L_{opt/UV}$} & 
    \colhead{$R_{\rm sub}$ (pc)} &
    \colhead{$3 \um$ Ratio} &
    \colhead{$6 \um$ Ratio} & 
    \colhead{$10 \um$ Ratio} &
    \colhead{$50 \um$ Ratio}
  }
  \startdata
  $\alpha_{ox}=1.1$ & 101.9\% & 1.78  & 1.49 & 1.55 & 1.54 & 1.14 \\
  $\alpha_{ox}=1.6$ & 69.7\%  & 1.57  & 1.00 & 1.00 & 1.00 & 1.00 \\
  $\alpha_{ox}=2.1$ & 84.9\%  & 1.54  & 1.56 & 1.34 & 1.18 & 1.09 \\
  \enddata
  \tablecomments{The ratio $L_{IR}/L_{opt/UV}$ is the amount of power
    emitted in the IR from $2 - 100 \um$ by a certain model as
    compared to the amount of power emitted in the optical/UV (from $2
    \um - 1000$ \AA) by the composite continuum by
    \cite{Richards2006}, when both are normalized at $L_{5100}$. In
    columns 4 through 7, all models are compared to the model with
    $\alpha_{ox} = 1.6$, $i=60^{\circ}$, $N_{H,0} = 10^{25}$
    cm$^{-2}$, $L/L_{Edd} = 0.1$, and $M_{BH}$ = 10$^{8} \msun$. }
\end{deluxetable}

\begin{deluxetable}{lrccccc}
  \tablewidth{0pt}
  \tablecaption{Variations in $\alpha_{ox}$ for $L/L_{Edd} = 0.01$, and $M_{BH} = 5 \times 10^{8} \msun$\label{aox_Ledd_0p01}}
  % \tablecolumns{9}
  \tablehead{
    \colhead{Parameter Change} &
    \colhead{$L_{IR}/L_{opt/UV}$} & 
    \colhead{$R_{\rm sub}$ (pc)} &
    \colhead{$3 \um$ Ratio} &
    \colhead{$6 \um$ Ratio} & 
    \colhead{$10 \um$ Ratio} &
    \colhead{$50 \um$ Ratio} 
  }
  \startdata
  $\alpha_{ox}=1.1$ & 123.8\% & 0.58  & 1.75 & 1.63 & 1.50 & 1.27 \\
  $\alpha_{ox}=1.6$ & 83.3\%  & 0.49  & 1.00 & 1.00 & 1.00 & 1.00 \\
  $\alpha_{ox}=2.1$ & 88.0\%  & 0.50  & 1.20 & 1.13 & 1.04 & 1.00 \\
  \enddata
  \tablecomments{The ratio $L_{IR}/L_{opt/UV}$ is the amount of power
    emitted in the IR from $2 - 100 \um$ by a certain model as
    compared to the amount of power emitted in the optical/UV (from $2
    \um - 1000$ \AA) by the composite continuum by
    \cite{Richards2006}, when both are normalized at $L_{5100}$.  For
    columns 4 through 7, all models are compared to the model with
    $\alpha_{ox} = 1.6$, $i=60^{\circ}$, $N_{H,0} = 10^{25}$
    cm$^{-2}$, $L/L_{Edd} = 0.01$, and $M_{BH} = 5 \times 10^{8}
    \msun$. }
\end{deluxetable}

\begin{deluxetable}{llllll}
  \tablewidth{0pt}
  \tablecaption{Variations in $M_{BH}$ \label{mbh_table}}
  % \tablecolumns{9}
  \tablehead{
    \colhead{Parameter Change} &
    \colhead{$L_{IR}/L_{opt/UV}$} & 
    \colhead{$3 \um$ Ratio} &
    \colhead{$6 \um$ Ratio} & 
    \colhead{$10 \um$ Ratio} &
    \colhead{$50 \um$ Ratio} 
  }
  \startdata
  $M_{BH} = 10^{8} \msun$          & 97.5\% & 1.00 & 1.00  & 1.00  & 1.00 \\
  $M_{BH} = 5 \times 10^{8} \msun$ & 83.3\% & 3.23 & 4.71  & 6.13  & 1.94 \\
  $M_{BH} = 10^{9} \msun$          & 82.6\% & 7.46 & 10.35 & 12.55 & 2.96 \\
  \enddata
  \tablecomments{All models are compared to the model with $M_{BH}$ = 10$^{8}
    \msun$, $\alpha_{ox} = 1.6$, $i=60^{\circ}$, $N_{H,0} = 10^{25}$
    cm$^{-2}$, and $L/L_{Edd} = 0.01$. The ratio $L_{IR}/L_{opt/UV}$ is the
    amount of power emitted in the IR from $2 - 100 \um$ by a certain model as
    compared to the amount of power emitted in the optical/UV (from $2 \um -
    1000$ \AA) by the composite continuum by \cite{Richards2006}, when both
    are normalized at $L_{5100}$.}
\end{deluxetable}

\begin{deluxetable}{llllll}
  \tablewidth{0pt}
  \tablecaption{Variations in $L/L_{Edd}$ \label{Ledd_table}}
  % \tablecolumns{9}
  \tablehead{
    \colhead{Parameter Change} &
    \colhead{$L_{IR}/L_{opt/UV}$} & 
    \colhead{$3 \um$ Ratio} &
    \colhead{$6 \um$ Ratio} & 
    \colhead{$10 \um$ Ratio} &
    \colhead{$50 \um$ Ratio} 
  }
  \startdata
  $L/L_{Edd} = 0.1$  & 69.7\% & 1.00 & 1.00 & 1.00  & 1.00 \\
  $L/L_{Edd} = 0.07$ & 79.6\% & 0.80 & 0.79 & 0.78  & 0.84 \\
  $L/L_{Edd} = 0.03$ & 87.9\% & 0.36 & 0.35 & 0.35  & 0.51 \\
  $L/L_{Edd} = 0.01$ & 97.5\% & 0.16 & 0.12 & 0.096 & 0.36 \\
  \enddata
  \tablecomments{All models are compared to the model with $L/L_{Edd} = 0.1$,
    $M_{BH} = 10^{8} \msun$, $\alpha_{ox} = 1.6$, $i=60^{\circ}$, and $N_{H,0}
    = 10^{25}$ cm$^{-2}$. The ratio $L_{IR}/L_{opt/UV}$ is the amount of power
    emitted in the IR from $2 - 100 \um$ by a certain model as compared to the
    amount of power emitted in the optical/UV (from $2 \um - 1000$ \AA) by the
    composite continuum by \cite{Richards2006}, when both are normalized at
    $L_{5100}$.}
\end{deluxetable}

\clearpage
%\input{figs}
%%%%%%%%%%%%%%%%%%%%%% 
% WIND STRUCTURE FIG
%%%%%%%%%%%%%%%%%%%%%% 
%%%%%%%%%%%%%%%%%%%%%%%%%%%%%%%%%%%%%%%%%%%%%%% 
% STREAMLINES
%%%%%%%%%%%%%%%%%%%%%%%%%%%%%%%%%%%%%%%%%%%%%%% 
%%%%%%%%%%%%%%%%%%%%%% 
% ALPHAOX FIGURE
%%%%%%%%%%%%%%%%%%%%%% 
%%%%%%%%%%%%%%%%%%%%%%
% MC3D GEOMETRY FIG
%%%%%%%%%%%%%%%%%%%%%% 
%%%%%%%%%%%%%%%%%%%%%% 
% MC3D TESTING FIGURE
%%%%%%%%%%%%%%%%%%%%%% 
%%%%%%%%%%%%%%%%%%%%%%%%%%%%%%%%%%%%%%%%%%%%%%% 
% COMPARE INCLINATION ANGLE, NH = 10^24
%%%%%%%%%%%%%%%%%%%%%%%%%%%%%%%%%%%%%%%%%%%%%%% 
%%%%%%%%%%%%%%%%%%%%%%%%%%%%%%%%%%%%%%%%%%%%%%% 
% COMPARE INCLINATION ANGLE, NH = 10^25
%%%%%%%%%%%%%%%%%%%%%%%%%%%%%%%%%%%%%%%%%%%%%%% 
%%%%%%%%%%%%%%%%%%%%%%%%%%%%%%%%%%%%%%%%%%%%%%% 
% FIDUCIAL PLOT
%%%%%%%%%%%%%%%%%%%%%%%%%%%%%%%%%%%%%%%%%%%%%%% 

%%%%%%%%%%%%%%%%%%%%%%%%%%%%%%%%%%%%%%%%%%%%%%% 
% COMPARE ALPHA_OX FIGURE L/Ledd = 0.1
%%%%%%%%%%%%%%%%%%%%%%%%%%%%%%%%%%%%%%%%%%%%%%% 
%%%%%%%%%%%%%%%%%%%%%%%%%%%%%%%%%%%%%%%%%%%%%%% 
% COMPARE ALPHA_OX FIGURE L/Ledd = 0.01, Mbh = 10^8
%%%%%%%%%%%%%%%%%%%%%%%%%%%%%%%%%%%%%%%%%%%%%%% 
%%%%%%%%%%%%%%%%%%%%%%%%%%%%%%%%%%%%%%%%%%%%%%% 
% COMPARE MBH FIGURE
%%%%%%%%%%%%%%%%%%%%%%%%%%%%%%%%%%%%%%%%%%%%%%% 
%%%%%%%%%%%%%%%%%%%%%%%%%%%%%%%%%%%%%%%%%%%%%%% 
% COMPARE L/L_EDD FIGURE
%%%%%%%%%%%%%%%%%%%%%%%%%%%%%%%%%%%%%%%%%%%%%%% 
%%%%%%%%%%%%%%%%%%%%%%%%%%%%%%%%%%%%%%%%%%%%%%% 
% DISTINGUISH FIGURE
%%%%%%%%%%%%%%%%%%%%%%%%%%%%%%%%%%%%%%%%%%%%%%% 
%%%%%%%%%%%%%%%%%%%%%%%%%%%%%%%%%%%%%%%%%%%%%%% 
% Hot dust
%%%%%%%%%%%%%%%%%%%%%%%%%%%%%%%%%%%%%%%%%%%%%%% 
%%%%%%%%%%%%%%%%%%%%%%%%%%%%%%%%%%%%%%%%%%%%%%% 
\clearpage
%\bibliography{agn}

\begin{thebibliography}{63}
\expandafter\ifx\csname natexlab\endcsname\relax\def\natexlab#1{#1}\fi

\bibitem[{{Antonucci}(1993)}]{Antonucci1993}
{Antonucci}, R. 1993, ARA$\&$A, 31, 473

\bibitem[{{Antonucci} \& {Miller}(1985)}]{Antonucci1985}
{Antonucci}, R.~R.~J., \& {Miller}, J.~S. 1985, \apj, 297, 621

\bibitem[{{Barvainis}(1990)}]{Barvainis1990}
{Barvainis}, R. 1990, \apj, 353, 419

\bibitem[{{Beckert} \& {Duschl}(2004)}]{Beckert2004}
{Beckert}, T., \& {Duschl}, W.~J. 2004, \aap, 426, 445

\bibitem[{{Blandford} \& {Payne}(1982)}]{Blandford1982}
{Blandford}, R.~D., \& {Payne}, D.~G. 1982, \mnras, 199, 883

\bibitem[{{Chartas} {et~al.}(2003){Chartas}, {Brandt}, \&
  {Gallagher}}]{Chartas2003}
{Chartas}, G., {Brandt}, W.~N., \& {Gallagher}, S.~C. 2003, \apj, 595, 85

\bibitem[{{Crenshaw} {et~al.}(2003){Crenshaw}, {Kraemer}, \&
  {George}}]{Crenshaw2003}
{Crenshaw}, D.~M., {Kraemer}, S.~B., \& {George}, I.~M. 2003, \araa, 41, 117

\bibitem[{{Deo} {et~al.}(2011){Deo}, {Richards}, {Nikutta}, {Elitzur},
  {Gallagher}, {Ivezi{\'c}}, \& {Hines}}]{Deo2011}
{Deo}, R.~P., {Richards}, G.~T., {Nikutta}, R., {et~al.} 2011, \apj, 729, 108

\bibitem[{{Dopita} {et~al.}(1998){Dopita}, {Heisler}, {Lumsden}, \&
  {Bailey}}]{Dopita1998}
{Dopita}, M.~A., {Heisler}, C., {Lumsden}, S., \& {Bailey}, J. 1998, \apj, 498,
  570

\bibitem[{{Dorodnitsyn} {et~al.}(2011{\natexlab{a}}){Dorodnitsyn},
  {Bisnovatyi-Kogan}, \& {Kallman}}]{Dorod2011a}
{Dorodnitsyn}, A., {Bisnovatyi-Kogan}, G.~S., \& {Kallman}, T.
  2011{\natexlab{a}}, \apj, 741, 29

\bibitem[{{Dorodnitsyn} {et~al.}(2011{\natexlab{b}}){Dorodnitsyn}, {Kallman},
  \& {Bisnovatyi-Kogan}}]{Dorod2011b}
{Dorodnitsyn}, A., {Kallman}, T., \& {Bisnovatyi-Kogan}, G.~S.
  2011{\natexlab{b}}, ArXiv e-prints

\bibitem[{{Draine}(2003)}]{Draine2003}
{Draine}, B.~T. 2003, \araa, 41, 241

\bibitem[{{Dullemond} \& {van Bemmel}(2005)}]{Dullemond2005}
{Dullemond}, C.~P., \& {van Bemmel}, I.~M. 2005, \aap, 436, 47

\bibitem[{{Elitzur} \& {Shlosman}(2006)}]{Elitzur2006}
{Elitzur}, M., \& {Shlosman}, I. 2006, \apjl, 648, L101

\bibitem[{{Elvis} {et~al.}(1994){Elvis}, {Wilkes}, {McDowell}, {Green},
  {Bechtold}, {Willner}, {Oey}, {Polomski}, \& {Cutri}}]{Elvis1994}
{Elvis}, M., {Wilkes}, B.~J., {McDowell}, J.~C., {et~al.} 1994, \apjs, 95, 1

\bibitem[{{Everett}(2005)}]{Everett2005}
{Everett}, J.~E. 2005, ApJ, 631, 689

\bibitem[{{Everett} {et~al.}(2009){Everett}, {Gallagher}, \&
  {Keating}}]{Everett2009}
{Everett}, J.~E., {Gallagher}, S.~C., \& {Keating}, S.~K. 2009, in American
  Institute of Physics Conference Proceedings, Vol. 1201, {The Monster's Fiery
  Breath: Feedback in Galaxies, Groups, and Clusters}, ed. {S.~Heinz \&
  E.~Wilcots}, 56--59

\bibitem[{{Ferland} {et~al.}(1998){Ferland}, {Korista}, {Verner}, {Ferguson},
  {Kingdon}, \& {Verner}}]{Ferland1998}
{Ferland}, G.~J., {Korista}, K.~T., {Verner}, D.~A., {et~al.} 1998, PASP, 110,
  761

\bibitem[{{Fritz} {et~al.}(2006){Fritz}, {Franceschini}, \&
  {Hatziminaoglou}}]{Fritz2006}
{Fritz}, J., {Franceschini}, A., \& {Hatziminaoglou}, E. 2006, \mnras, 366, 767

\bibitem[{{Gallagher} {et~al.}(2006){Gallagher}, {Brandt}, {Chartas},
  {Priddey}, {Garmire}, \& {Sambruna}}]{Gallagher2006}
{Gallagher}, S.~C., {Brandt}, W.~N., {Chartas}, G., {et~al.} 2006, \apj, 644,
  709

\bibitem[{{Gibson} {et~al.}(2009){Gibson}, {Jiang}, {Brandt}, {Hall}, {Shen},
  {Wu}, {Anderson}, {Schneider}, {Vanden Berk}, {Gallagher}, {Fan}, \&
  {York}}]{Gibson2009}
{Gibson}, R.~R., {Jiang}, L., {Brandt}, W.~N., {et~al.} 2009, \apj, 692, 758

\bibitem[{{Grimes} {et~al.}(2004){Grimes}, {Rawlings}, \&
  {Willott}}]{Grimes2004}
{Grimes}, J.~A., {Rawlings}, S., \& {Willott}, C.~J. 2004, \mnras, 349, 503

\bibitem[{{Hao} {et~al.}(2007){Hao}, {Weedman}, {Spoon}, {Marshall},
  {Levenson}, {Elitzur}, \& {Houck}}]{Hao2007}
{Hao}, L., {Weedman}, D.~W., {Spoon}, H.~W.~W., {et~al.} 2007, \apjl, 655, L77

\bibitem[{{Hao} {et~al.}(2005{\natexlab{a}}){Hao}, {Strauss}, {Fan},
  {Tremonti}, {Schlegel}, {Heckman}, {Kauffmann}, {Blanton}, {Gunn}, {Hall},
  {Ivezi{\'c}}, {Knapp}, {Krolik}, {Lupton}, {Richards}, {Schneider},
  {Strateva}, {Zakamska}, {Brinkmann}, \& {Szokoly}}]{Hao2005b}
{Hao}, L., {Strauss}, M.~A., {Fan}, X., {et~al.} 2005{\natexlab{a}}, \aj, 129,
  1795

\bibitem[{{Hao} {et~al.}(2005{\natexlab{b}}){Hao}, {Spoon}, {Sloan},
  {Marshall}, {Armus}, {Tielens}, {Sargent}, {van Bemmel}, {Charmandaris},
  {Weedman}, \& {Houck}}]{Hao2005}
{Hao}, L., {Spoon}, H.~W.~W., {Sloan}, G.~C., {et~al.} 2005{\natexlab{b}},
  \apjl, 625, L75

\bibitem[{{Hawley}(2011)}]{Hawley2011}
{Hawley}, J.~F. 2011, in IAU Symposium, Vol. 275, IAU Symposium, ed.
  {G.~E.~Romero, R.~A.~Sunyaev, \& T.~Belloni}, 50--58

\bibitem[{{Hill} {et~al.}(1996){Hill}, {Goodrich}, \& {Depoy}}]{Hill1996}
{Hill}, G.~J., {Goodrich}, R.~W., \& {Depoy}, D.~L. 1996, \apj, 462, 163

\bibitem[{{H{\"o}nig} \& {Kishimoto}(2010)}]{Honig2010}
{H{\"o}nig}, S.~F., \& {Kishimoto}, M. 2010, \aap, 523, A27+

\bibitem[{{Hopkins} {et~al.}(2011){Hopkins}, {Hayward}, {Narayanan}, \&
  {Hernquist}}]{HopkinsEtAl2011}
{Hopkins}, P.~F., {Hayward}, C.~C., {Narayanan}, D., \& {Hernquist}, L. 2011,
  \mnras, 2115

\bibitem[{{Jaffe} {et~al.}(2004){Jaffe}, {Meisenheimer}, {R{\"o}ttgering},
  {Leinert}, {Richichi}, {Chesneau}, {Fraix-Burnet}, {Glazenborg-Kluttig},
  {Granato}, {Graser}, {Heijligers}, {K{\"o}hler}, {Malbet}, {Miley},
  {Paresce}, {Pel}, {Perrin}, {Przygodda}, {Schoeller}, {Sol}, {Waters},
  {Weigelt}, {Woillez}, \& {de Zeeuw}}]{Jaffe2004}
{Jaffe}, W., {Meisenheimer}, K., {R{\"o}ttgering}, H.~J.~A., {et~al.} 2004,
  \nat, 429, 47

\bibitem[{{Just} {et~al.}(2007){Just}, {Brandt}, {Shemmer}, {Steffen},
  {Schneider}, {Chartas}, \& {Garmire}}]{Just2007}
{Just}, D.~W., {Brandt}, W.~N., {Shemmer}, O., {et~al.} 2007, \apj, 665, 1004

\bibitem[{{K\"{o}nigl} \& {Kartje}(1994)}]{Konigl1994}
{K\"{o}nigl}, A., \& {Kartje}, J.~F. 1994, ApJ, 434, 446

\bibitem[{{Krolik}(2007)}]{Krolik2007}
{Krolik}, J.~H. 2007, \apj, 661, 52

\bibitem[{{Krolik} \& {Begelman}(1988)}]{Krolik1988}
{Krolik}, J.~H., \& {Begelman}, M.~C. 1988, \apj, 329, 702

\bibitem[{{Lawrence}(1991)}]{Lawrence1991}
{Lawrence}, A. 1991, \mnras, 252, 586

\bibitem[{{Lawrence} \& {Elvis}(2010)}]{Lawrence2010}
{Lawrence}, A., \& {Elvis}, M. 2010, \apj, 714, 561

\bibitem[{{Mason} {et~al.}(2009){Mason}, {Levenson}, {Shi}, {Packham},
  {Gorjian}, {Cleary}, {Rhee}, \& {Werner}}]{Mason2009}
{Mason}, R.~E., {Levenson}, N.~A., {Shi}, Y., {et~al.} 2009, \apjl, 693, L136

\bibitem[{{Mathis} {et~al.}(1977){Mathis}, {Rumpl}, \&
  {Nordsieck}}]{Mathis1977}
{Mathis}, J.~S., {Rumpl}, W., \& {Nordsieck}, K.~H. 1977, \apj, 217, 425

\bibitem[{{Murray} {et~al.}(1995){Murray}, {Chiang}, {Grossman}, \&
  {Voit}}]{MCGV1995}
{Murray}, N., {Chiang}, J., {Grossman}, S.~A., \& {Voit}, G.~M. 1995, \apj,
  451, 498

\bibitem[{{Nenkova} {et~al.}(2008{\natexlab{a}}){Nenkova}, {Sirocky},
  {Ivezi{\'c}}, \& {Elitzur}}]{Nenkova2008a}
{Nenkova}, M., {Sirocky}, M.~M., {Ivezi{\'c}}, {\v Z}., \& {Elitzur}, M.
  2008{\natexlab{a}}, \apj, 685, 147

\bibitem[{{Nenkova} {et~al.}(2008{\natexlab{b}}){Nenkova}, {Sirocky},
  {Nikutta}, {Ivezi{\'c}}, \& {Elitzur}}]{Nenkova2008b}
{Nenkova}, M., {Sirocky}, M.~M., {Nikutta}, R., {Ivezi{\'c}}, {\v Z}., \&
  {Elitzur}, M. 2008{\natexlab{b}}, \apj, 685, 160

\bibitem[{{Netzer} {et~al.}(2007){Netzer}, {Lutz}, {Schweitzer}, {Contursi},
  {Sturm}, {Tacconi}, {Veilleux}, {Kim}, {Rupke}, {Baker}, {Dasyra},
  {Mazzarella}, \& {Lord}}]{Netzer2007}
{Netzer}, H., {Lutz}, D., {Schweitzer}, M., {et~al.} 2007, \apj, 666, 806

\bibitem[{{Pier} \& {Krolik}(1992)}]{Pier1992}
{Pier}, E.~A., \& {Krolik}, J.~H. 1992, \apj, 401, 99

\bibitem[{{Proga} \& {Kallman}(2004)}]{Proga2004}
{Proga}, D., \& {Kallman}, T.~R. 2004, \apj, 616, 688

\bibitem[{{Reichard} {et~al.}(2003){Reichard}, {Richards}, {Hall}, {Schneider},
  {Vanden Berk}, {Fan}, {York}, {Knapp}, \& {Brinkmann}}]{Reichard2003}
{Reichard}, T.~A., {Richards}, G.~T., {Hall}, P.~B., {et~al.} 2003, \aj, 126,
  2594

\bibitem[{{Richards} {et~al.}(2006){Richards}, {Lacy}, {Storrie-Lombardi},
  {Hall}, {Gallagher}, {Hines}, {Fan}, {Papovich}, {Vanden Berk}, {Trammell},
  {Schneider}, {Vestergaard}, {York}, {Jester}, {Anderson}, {Budav{\'a}ri}, \&
  {Szalay}}]{Richards2006}
{Richards}, G.~T., {Lacy}, M., {Storrie-Lombardi}, L.~J., {et~al.} 2006, ApJS,
  166, 470

\bibitem[{{Sanders} {et~al.}(1989){Sanders}, {Phinney}, {Neugebauer}, {Soifer},
  \& {Matthews}}]{Sanders1989}
{Sanders}, D.~B., {Phinney}, E.~S., {Neugebauer}, G., {Soifer}, B.~T., \&
  {Matthews}, K. 1989, ApJ, 347, 29

\bibitem[{{Schartmann} {et~al.}(2011){Schartmann}, {Krause}, \&
  {Burkert}}]{Schartmann2011}
{Schartmann}, M., {Krause}, M., \& {Burkert}, A. 2011, \mnras, 415, 741

\bibitem[{{Schartmann} {et~al.}(2005){Schartmann}, {Meisenheimer}, {Camenzind},
  {Wolf}, \& {Henning}}]{Schartmann2005}
{Schartmann}, M., {Meisenheimer}, K., {Camenzind}, M., {Wolf}, S., \&
  {Henning}, T. 2005, \aap, 437, 861

\bibitem[{{Schartmann} {et~al.}(2008){Schartmann}, {Meisenheimer}, {Camenzind},
  {Wolf}, {Tristram}, \& {Henning}}]{Schartmann2008}
{Schartmann}, M., {Meisenheimer}, K., {Camenzind}, M., {et~al.} 2008, \aap,
  482, 67

\bibitem[{{Schweitzer} {et~al.}(2006){Schweitzer}, {Lutz}, {Sturm}, {Contursi},
  {Tacconi}, {Lehnert}, {Dasyra}, {Genzel}, {Veilleux}, {Rupke}, {Kim},
  {Baker}, {Netzer}, {Sternberg}, {Mazzarella}, \& {Lord}}]{Schweitzer2006}
{Schweitzer}, M., {Lutz}, D., {Sturm}, E., {et~al.} 2006, ApJ, 649, 79

\bibitem[{{Shi} \& {Krolik}(2008)}]{Shi2008}
{Shi}, J., \& {Krolik}, J.~H. 2008, \apj, 679, 1018

\bibitem[{{Siebenmorgen} {et~al.}(2005){Siebenmorgen}, {Haas}, {Kr{\"u}gel}, \&
  {Schulz}}]{Siebenmorgen2005}
{Siebenmorgen}, R., {Haas}, M., {Kr{\"u}gel}, E., \& {Schulz}, B. 2005, \aap,
  436, L5

\bibitem[{{Simpson}(2005)}]{Simpson2005}
{Simpson}, C. 2005, \mnras, 360, 565

\bibitem[{{Simpson} \& {Rawlings}(2000)}]{Simpson2000}
{Simpson}, C., \& {Rawlings}, S. 2000, \mnras, 317, 1023

\bibitem[{{Steffen} {et~al.}(2003){Steffen}, {Barger}, {Cowie}, {Mushotzky}, \&
  {Yang}}]{Steffen2003}
{Steffen}, A.~T., {Barger}, A.~J., {Cowie}, L.~L., {Mushotzky}, R.~F., \&
  {Yang}, Y. 2003, \apjl, 596, L23

\bibitem[{{Steffen} {et~al.}(2006){Steffen}, {Strateva}, {Brandt}, {Alexander},
  {Koekemoer}, {Lehmer}, {Schneider}, \& {Vignali}}]{Steffen2006}
{Steffen}, A.~T., {Strateva}, I., {Brandt}, W.~N., {et~al.} 2006, AJ, 131, 2826

\bibitem[{{Tananbaum} {et~al.}(1979){Tananbaum}, {Avni}, {Branduardi}, {Elvis},
  {Fabbiano}, {Feigelson}, {Giacconi}, {Henry}, {Pye}, {Soltan}, \&
  {Zamorani}}]{Tananbaum1979}
{Tananbaum}, H., {Avni}, Y., {Branduardi}, G., {et~al.} 1979, ApJ, 234, L9

\bibitem[{{Tristram} {et~al.}(2007){Tristram}, {Meisenheimer}, {Jaffe},
  {Schartmann}, {Rix}, {Leinert}, {Morel}, {Wittkowski}, {R{\"o}ttgering},
  {Perrin}, {Lopez}, {Raban}, {Cotton}, {Graser}, {Paresce}, \&
  {Henning}}]{Tristram2007}
{Tristram}, K.~R.~W., {Meisenheimer}, K., {Jaffe}, W., {et~al.} 2007, \aap,
  474, 837

\bibitem[{{Urry} \& {Padovani}(1995)}]{Urry1995}
{Urry}, C.~M., \& {Padovani}, P. 1995, PASP, 107, 803

\bibitem[{{van Hoof} {et~al.}(2004){van Hoof}, {Weingartner}, {Martin}, {Volk},
  \& {Ferland}}]{vanHoof2004}
{van Hoof}, P.~A.~M., {Weingartner}, J.~C., {Martin}, P.~G., {Volk}, K., \&
  {Ferland}, G.~J. 2004, \mnras, 350, 1330

\bibitem[{{Vollmer} {et~al.}(2004){Vollmer}, {Beckert}, \&
  {Duschl}}]{Vollmer2004}
{Vollmer}, B., {Beckert}, T., \& {Duschl}, W.~J. 2004, \aap, 413, 949

\bibitem[{{Wolf}(2003)}]{Wolf2003}
{Wolf}, S. 2003, Computer Physics Communications, 150, 99

\end{thebibliography}

\end{document}